\documentclass[prc,notitlepage,superscriptaddress,nofootinbib,showkeys,showpacs,reprint]{revtex4-1}
\usepackage[english]{babel}
\usepackage{amsfonts}
\usepackage{amsmath}
\usepackage{amssymb}
\usepackage{euscript}
\usepackage{graphicx}
\usepackage{mathrsfs}

\def\magic#1{#1$^{b)}$}

\newcommand{\lam}{\lambdabar}
\newcommand{\eus}{\EuScript}
\newcommand{\eps}{\varepsilon}

\newcommand{\ta}{\vartheta}

\newcommand{\ffi}{\varphi}

\newcommand{\ra}{\rangle}
\newcommand{\la}{\langle}

\newcommand{\lsim}{\stackrel{<}{_\sim}}
\newcommand{\g}{\gamma}
\newcommand{\om}{\hbar\omega}

\newcommand{\al}{\alpha}
\newcommand{\bt}{\beta}
\newcommand{\gm}{\gamma}

\newcommand{\dl}{\delta}

\newcommand{\lm}{\lambda}

\newcommand{\HH}{\eus H}

\newcommand{\nbl}{\nabla}
\newcommand{\pt}{\partial}
\newcommand{\sg}{\sigma}
\newcommand{\DD}{\Delta}

\begin{document}
\title{Description of the shape of medium and heavy nuclei using a finite deformed one-particle potential with deformation-dependent diffuseness}

\author{G.~I.~Bykhalo}

\affiliation{Lomonosov Moscow State University, Department of Physics, Moscow, 119991 Russia}


\author{V.~N.~Orlin}

\affiliation{Lomonosov Moscow State University, Skobeltsyn Institute of Nuclear Physics, Moscow, 119991 Russia}

\author{K.~A.~Stopani}

\affiliation{Lomonosov Moscow State University, Skobeltsyn Institute of Nuclear Physics, Moscow, 119991 Russia}
\email{hatta@depni.sinp.msu.ru}

\date{\today}

\begin{abstract}
Adjustment of the behavior of the potential energy of nuclear deformation, defined as the sum of the energies of lowest-lying occupied single-particle levels in a deformed finite potential with a pairing correction, is considered  by taking into account the dependence of the diffuseness of the surface of the potential on deformation. To verify this approach we construct a non-axial ellipsoidally-deformed potential with a slight (of the order of 1\%) variation of the surface diffuseness at moderate deformations. Parameters of the variation are determined from a calculation for a group of 36 deformed, transitional, and spherical nuclei in the range $50 \le A \le 241$ using existing experimental data on static quadrupole moments. In order to test the performed parameter choice, we applied the model to description of the ground-state deformation of the isotopic chains $^{95-132}$Cd and $^{74-106}$Sr. The obtained description of the ground state deformation of the considered nuclei is in agreement with experimental data and results of macro-microscopic models and calculations using the Hartree-Fock-Bogoliubov method.
\end{abstract}


\maketitle

\section*{Introduction}

Accurate description of the ground-state shape of nuclei is important in studies of nuclear reactions with excitation of low-lying single-particle and collective degrees of freedom, photonuclear reactions in the giant dipole resonance formation range ($E_\gamma < 30$~MeV), and other fields. A number of general approaches to this problem have been developed over time, including such advanced as the macro-microscopic and self-consistent mean field models as well as, for example, models which attracted attention in the last decade to the question of dominance of prolate shapes of nuclei in ground state \cite{hamamoto-2009,takahara-2011,bonatsos-2017}.

The modern macro-microscopic method of calculation of the internal energy of nucleus, such as the Finite Range Droplet Model (FRDM) \cite{moeller-1995,moeller-2016,moeller-2012} or the nuclear mass model of Wang \textit{et al.} \cite{wang-2010,wang-2010b} where the macroscopic energy term is calculated using a variant of the liquid-drop formula, while the microscopic term consists of the Strutinsky shell correction \cite{strutinsky-1967b,strutinsky-1968} or the Wigner-Kirkwood expansion of the one-body partition function \cite{bhagwat-2012} and the contribution of pairing interaction, has become a very efficient means of estimation of mass defects, binding energies, and deformations of nuclei in ground state and also of description of the fission process, including the hypothetical region of super-heavy nuclei \cite{johansson-1970,pauli-1973}. In Ref. \cite{moeller-1995,moeller-2016}, using the folded Yukawa potential as the phenomenological one-particle potential, the mass defects, binding energies, and ground state deformations for 9318 nuclei from $^{16}$O to $A=339$ were calculated. The calculation spans between the proton and neutron drip lines, and the r.m.s. errors of the calculated masses do not exceed about 0.56 MeV in comparison with the experimental values and are even less (0.35 MeV) for the range $N \ge 65$.

A significant theoretical interest is attracted to the methods of calculation of the internal energy of deformed nucleus based on the computation of self-consistent average nuclear field.  Usage of the Skyrme interaction \cite{vautherin-1969, vautherin-1972}, introduction of the Gogny forces \cite{gogny-1973}, and, finally, formulation of the relativistic mean field models \cite{walecka-1974, boguta-1977} resulted in development of three ``standard'' models of the self-consistent nuclear average field method that are in a widespread use today and fare comparably with the macro-microscopic method at the quantitative level.

Present day self-consistent mean-field calculations performed using the Skyrme force \cite{bender-2003,ryssens-2019,goriely-2013,bruslib} or the density-dependent effective Gogny interaction \cite{decharg-1980,hilaire-2007,amedee} or the relativstic mean field models \cite{serot-1984,reinhard-1989,serot-1992,ring-1996} always include the pairing interaction, typically described in terms of the Hartree--Fock--Bogoliubov theory (HFB) \cite{mang-1975, goodman-1979, blaizot-1986, ring-1980}. A number of large-scale calculations using the HFB method have been performed, where the ground-state values of the deformation parameters and nuclear masses were determined for isotopes between the proton and neutron drip-lines (about $7000$ nuclei in \cite{hilaire-2007}), the regions of super- and hyper-deformed nuclei were predicted \cite{hilaire-2007,ryssens-2019}, the factors affecting the accuracy of description of the fission barriers were studied \cite{ryssens-2019}. Data from such calculations are available on-line \cite{bruslib,amedee}. 

The purpose of the present work is construction of a simple phenomenological model describing the ground-state deformation of majority of nucleon-stable nuclei, including the nuclei outside of the $\beta$-stability valley, having as a rule a moderate value of quadrupole deformation ($-0.5 \le \beta_2 \le +0.5$). For this we modify the method of prediction of the deformation proposed by Nilsson \cite{nilsson-1955}, in which the deformation energy of nucleus is assumed to be proportional to the sum of single-particle energies of nucleons occupying the lowest levels in the deformed shell-model potential. Early calculations of this type in Refs. \cite{mottelson-1959,bes-1961,bohr-mottelson-1} allowed to reproduce experimental quadrupole moments in the rare-earth and actinide regions and demonstrated the importance of shell effects for estimation of deformation of nuclei. Later attempts of application of this method, however, showed that without modifications it is unable to correctly describe the equilibrium shape of the majority of lighter and less strongly deformed nuclei, since the total single-particle energy is not a correct description of the internal energy of nucleus \cite{brack-1992,brack-1972}.


In medium and heavy nuclei the thickness of diffuse layer is small in comparison with the radius. It has been shown by Myers and Swiatecki \cite{myers-1969,myers-1973} in terms of the liquid drop model that in the case of spherical shape the distribution of nucleon density in such systems can be written as $\rho(r) = \rho_0 f((r-R_0)/a_0)$, where $\rho_0$ is the density at the center, and $f$ denotes a Woods--Saxon-type form-factor with the radius of the surface $R_0$ and the diffuseness parameter $a_0$. Deformation (of ellipsoidal or other type) of such nucleus can be represented as a bodily displacement of the surface at each point  keeping the enclosed volume constant, which leads to replacement of the radius $R_0$ with $R_0(\vartheta,\varphi)$. At the same time angular dependence of the diffuseness parameter $a_0(\vartheta,\varphi)$ is introduced so that the gradient of density $\rho$ at each point of the surface is constant and does not depend on deformation \cite{bohr-mottelson-2}. If the deformation type is specified such transformations of the density distribution and the closely related one-particle potential are completely determined by their initial spherical forms.

In an actual nuclear system the requirement of independence of the surface gradient (and, consequently, the thickness of the diffuse layer) on the value of deformation is, probably, not met, due to a large effect of deformation on distribution of single-particle levels. The thickness of the diffuse layer, however, can be expected to be constant at each point of the surface. The obtained finite deformed potential is completely determined by its initial spherical form with the exception that the thickness of the diffuse layer may, within certain limits, vary at different values of the deformation, while remaining thin.

We use a finite deformed Woods--Saxon one-particle potential, the diffuseness of the surface of which is a function of deformation. It will be shown that this aspect plays a principal role in correction of the total single-particle energy, allowing the effect of residual forces to be approximately included in it.

As is well-known, the sum of single-particle energies $E$ in a finite deformed one-particle potential as a function of deformation (which is alternatively referred to throughout the text as potential energy of deformation) shows a large-scale oscillating structure brought about by the shell-model effects. In particular, one of its local minima usually approximately corresponds to the equilibrium ground-state shape of the nucleus, though it is often not the absolute minimum. The relative depth of different minima of $E$ is to a large extent determined by two competing factors: the rate of growth of the total energy of the most energetic nucleons concentrated near the surface, and the rate of decrease of the energy of interaction of protons with the average Coulomb field as the deformation increases. 

The surface energy of a system of nucleons in a finite one-particle potential at a given deformation is in direct relation with the thickness of its surface layer. Its increase results in increase of the surface energy, and decrease, in turn, results in decrease of the surface energy. From analysis of the experimental data on nucleon scattering off nuclei such as \cite{koning-2003} the surface diffuseness parameter $a$ can be estimated only for spherical nuclei, and its behavior is unknown as the nuclear shape undergoes deformation. It is usually assumed that it remains constant.

However, as noted above, this assumption is not undoubted. From the structure of levels in one-particle deformed potential it can be noticed that the density of occupied levels near the Fermi surface and their positions with respect to the unoccupied levels undergo noticeable changes as a response to deformation of the potential. It implies that the resulting diffuseness of the nuclear surface and, consequently, the diffuseness of the surface of a realistic one-particle potential should also demonstrate similar variations.

In nuclei with closed or almost closed outer shells at zero deformation and in stably deformed nuclei (where the amplitude of surface oscillations is small in comparison with the static deformation) at a certain non-zero deformation the density of occupied single-particle levels increases near the Fermi surface due to formation of a group of closely positioned levels which are distanced from the higher unoccupied orbits by an energy gap. In both cases one should expect reduction of the surface diffuseness in the region of such structures.


In the process of deformation of a nucleus with closed shells the internal energy of the system of nucleons increases simultaneously with fanning out of the energy levels, which leads to increase of the diffuseness of the surface of the nucleus and, therefore, of its surface energy. In such nuclei the equlibrium deformation has to be close to zero. On the other hand, in mid-shell nuclei the deformed shape can be more energetically beneficial since, as the system approaches the region with more dense spacing of the levels near the Fermi surface, the diffuseness of the surface decreases, leading to decrease of the surface and, consequently, total internal energies of the nucleus.


This connection of variation of diffuseness of nuclear surface with deformation will be demonstrated in Section \ref{section3}.

\section{Ellipsoidal one-particle potential with a deformation-dependent diffuseness of surface}

Throughout the text we neglect neutron-proton correlations assuming that neutrons and protons perform independent motion in separate potentials.

In this work, similarly to \cite{ishkhanov-2005}, we construct the anisotropic shell-model potential well $V(r,\ta,\ffi)$ by ellipsoidally deforming the real part of the global spherical optical potential \cite{koning-2003} evaluated at the energy $\varepsilon=E_F$ (where $\varepsilon$ is the energy of the nucleon, $E_F$ is the Fermi energy), which we treat as a spherical one-paticle potential $V(r) \equiv \left. V(r,\ta,\ffi) \right|_{\beta=0,\gamma=0}$, where $\beta$ is the axial deformation parameter and $\gamma$ is the nonaxiality parameter of the ellipsoid. Such choice should present a reliable estimation of the parameters of the potential $V(r)$, since the parameterization of the global optical potential had been performed by its authors on the basis of a large amount of experimental data on nucleon scattering on, mostly, spherical nuclei. As a result we begin with the following expression for the spherical potential:
\begin{equation}
V(r) = -U_\text{nucl}f_\text{nucl}(r) + 2\lam_\pi^2 U_\text{ls}\frac{1}{r}
\frac{df_\text{ls}(r)}{dr}\,{\mathbf{l}\mathbf{s}} + V_\text{Coul}(r),
\label{1}
\end{equation}
where the three terms correspond, respectively, to the contributions of the nuclear, spin-orbit, and Coulomb interaction,
\begin{eqnarray}
U_\text{nucl}=\left(59.3+(-1)^{1-q}21\frac{N-Z}{A}-0.024 A\right)\times\nonumber\\
\left(1+q(0.007067+4.23\times10^{-6}A)\frac{1.73 Z}{R_\text{Coul}}\right) \, [\text{MeV}]
\label{2}
\end{eqnarray}
is the depth of the nuclear potential well, $q$ is the charge number of the nucleon, equal to 0 for a neutron, and 1 for proton, $N$ and $Z$ are the neutron and proton numbers of the nucleus, $A=N+Z$ is the mass number, $R_\text{Coul}=1.198 A^{1/3}+0.697 A^{-1/3}+12.994 A^{-4/3}$ [fm] is the Coulomb radius of the nucleus,
\begin{equation}
U_\text{ls} = 5.922 + 0.003 A \quad [\text{MeV}]
\label{3}
\end{equation}
is the depth of the spin-orbit potential well, $\lam_\pi = 1.414$ MeV is the Compton wavelength of pion,
\begin{equation}
f_i(r) = \frac{1}{1+\exp[(r-R_i)/a_i]},
\label{4}
\end{equation}
are the radial Woods--Saxon form factors for the nuclear (where $i=\text{``nucl''}$) and spin-orbit (where $i=\text{``ls''}$) interaction terms with the radii $R_\text{nucl}=1.3039 \times A^{1/3}-0.4054$ [fm], $R_\text{ls}=1.1854A^{1/3}
-0.647$ [fm], and the diffuseness parameters $a_\text{nucl}=0.6778-1.487\times10^{-4}A$ [fm], $a_\text{ls}=0.59$ [fm], respectively, and
\begin{equation}
V_\text{Coul}(r) = 
\begin{cases}
\frac{qZe^2}{2R_\text{Coul}}\left[3-(r/R_\text{Coul})^2\right] &
\text{for $r\le R_\text{Coul}$}, \\
\frac{qZe^2}{r} & \text{for $r>R_\text{Coul}$}
\end{cases}
\label{5}
\end{equation}
is the Coulomb potential.

The deformed axially-asymmetric one-paticle potential should also contain three components:
\begin{equation}
V(r,\ta,\ffi) = V_\text{nucl}(r,\ta,\ffi)+V_\text{ls}(r,\ta,\ffi)+
V_\text{coul}(r,\ta,\ffi),
\label{6}
\end{equation}
where $\{r,\ta,\ffi\}$ are the spherical coordinates of the nucleon;
\begin{equation}
V_\text{nucl}(r,\ta,\ffi)=-U_\text{nucl}f_\text{nucl}(r,\ta,\ffi)
\label{7}
\end{equation}
is the deformed nuclear Woods--Saxon potential;
\begin{equation}
V_\text{ls}(r,\ta,\ffi)=\lam_\pi^2U_\text{ls}(F+F^+),
\label{21}
\end{equation}
is the spin-orbit term of the deformed one-paticle potential described by an Hermitian operator,
\begin{equation}
F=(\nbl f_\text{ls}(r,\ta,\ffi)\times{\bf p}) \mathbf{s}
\label{22}
\end{equation}
is the operator of interaction of the spin of a moving nucleon with the form factor of the average field \cite{landau-1977},
$\mathbf{p} = -i\nabla$ is the nucleon's momentum,  $\mathbf{s}$ is its spin;
\begin{multline}
V_\text{Coul}(r,\ta,\ffi)=\frac{3}{4\pi}\,\frac{qZe^2}{R_\text{Coul}^3} \times \\
\int\limits_0^{2\pi}\!\!d\ffi'\int\limits_0^\pi\!\!\sin\ta'd\ta'
\!\!\int\limits_0^{R_\text{Coul}(\ta'\ffi')}\!\!\!\frac{(r')^2dr'}{\sqrt{r^2+(r')^2
-2rr'\cos\beta}}\,,
\label{29}
\end{multline}
is the Coulomb term of the one-particle potential, generalizing formula \eqref{5} to the case of a uniformly charged ellipsoid of the same volume with a surface $R_\text{Coul}(\ta,\ffi)$; $\cos\bt=\cos\ta\cos\ta'+\sin\ta\sin\ta'\cos(\ffi-\ffi')$ is the cosine of the angle between the radius vectors {\bf r} and {\bf r$'$}.

For medium and heavy nuclei (see discussion in Introduction) the anisotropic Woods--Saxon form factor can be represented in the form \cite{bohr-mottelson-2}:
\begin{equation}
f_i(r,\ta,\ffi) = \frac{1}{1+\exp\left(\frac{r-R_i(\ta,\ffi)}{a_i(\ta,\ffi)}\right)}, \quad i=\text{nucl},\text{ls}
\label{8}
\end{equation}
where $R_i(\ta,\ffi)$ and $a_i(\ta,\ffi)$ are, respectively, the radius and diffuseness of the surface at the direction $\{\ta,\ffi\}$, which satisfy the following requirements: (a) conservation the volume enclosed inside the $R_i(\ta,\ffi)$ surface compared to the volume of the non-defomed potential, (b) independence of the gradient of $V_i(r,\ta,\ffi)$ on $\ta$ and $\ffi$ (i.e., the thickness of the surface layer is constant at any point of the surface).

We limit ourselves to consideration of ellipsoidal deformations. In this case
\begin{equation}
R_i(\ta,\ffi)= R'_i\left(\frac{\sin^2
	\ta\cos^2\ffi}{b^2}+\frac{\sin^2\ta\sin^2\ffi}{d^2}+
\frac{\cos^2\ta}{c^2}\right)^{-\frac{1}{2}},
\label{9}
\end{equation}
where $b^2$, $d^2$, $c^2$ are defined by the relationships
\begin{eqnarray}
\label{10}
b^2 &=& 1 + \sqrt{\frac{5}{\pi}}\bt\cos\left(\gm-\frac{2\pi}{3}\right)\\ 
d^2 &=& 1 + \sqrt{\frac{5}{\pi}}\bt\cos\left(\gm+\frac{2\pi}{3}\right)\nonumber\\
c^2 &=& 1 + \sqrt{\frac{5}{\pi}}\bt\cos\gm;\nonumber
\end{eqnarray}
$\beta$ is the axial deformation parameter, $\gamma$ is the non-axiality parameter, $ R'_ib$, $ R'_ic$, $ R'_id$ are the semi-axis lengths, and the factor $ R'_i= R'_i(\bt,\gm)$ is chosen so as to keep constant the volume of the sphere undergoing deformation:
\begin{eqnarray}
(R'_i)^3bdc=R_i^3 \,.
\label{11}
\end{eqnarray}

The surface $R_\text{Coul}(\ta,\ffi)$ of a uniformly charged ellipsoid creating a Coulomb field \eqref{29} is determined by the same formulas 
\eqref{9}--\eqref{11}, with replacement of the radius $R_i$ with $R_\text{Coul}$.

The squared gradient of the spherical form factor at the surface is $\left(\frac{df_i(r)}{dr}\right)^2_{r=R_i}
=\frac{1}{16a_i^2}$. Therefore, we formulate the condition that the gradient is kept constant at the surface as follows:
\begin{equation}
[\nbl f_i(r,\ta,\ffi)]^2_{r=R_i(\ta,\ffi)}
=\frac{1}{16 (a'_i)^2}\,,
\label{12}
\end{equation}
where the parameter $ a'_i=a'_i(\bt,\g)$ and is equal to $a_i$ for $\bt=\g=0$. Calculating  in \eqref{12} the differential operator $\nbl$ in the local spherical basis centered at the considered point of the surface, we find that the diffuseness of the surface of the deformed potential is determined by the value
\begin{multline}
a_i(\ta,\ffi) =   a'_i\Biggl\{1+ \frac{1}{R_i^2(\ta,\ffi)}
\Biggl[\frac{\partial R_i(\ta,\ffi)}{\partial\ta}\Biggr]^2
+\\
\frac{1}{R_i^2(\ta,\ffi)\sin^2\ta}\Biggl[\frac{\partial
R_i(\ta,\ffi)}{\partial\ffi}\Biggr]^2\Biggr\}^\frac{1}{2}.
\label{13}
\end{multline}
The $a'_i(\bt,\gm)$ factor entering this expression will be parameterized in Section \ref{section3}.

The overall shape of an ellipsoid is defined by pair-wise ratios of its semi-axes (see \eqref{9},\eqref{10}). Each of the possible ellipsoidal shapes corresponds to 6 different points in the $(\bt,\gm)$ plane, differing only in redesignation of the intrinsic coordinate axes $(x, y, z)$ \cite{bohr-mottelson-2}. Therefore, the searches of the equilibrium nuclear shape in the ground state can be limited to computations only in the $0\le\gm\le\pi/3$ sector, containing every possible ellipsoidal shape. Specifically, the $\gamma=0$ value corresponds to prolate, and $\gamma=\pi/3$ to oblate axially-symmetric shapes with the symmetry axes, respectively, $z$ and $y$.

In local spherical coordinate frame centered at $(r,\ta,\ffi)$, the operator $F$ can be represented as
\begin{equation}
F=F^{(r)}+F^{(\ta)}+F^{(\ffi)},
\label{25}
\end{equation}
where
\begin{equation}
F^{(r)}=\frac{1}{r}\frac{\pt f_\text{ls}(r,\ta,\ffi)}{\pt r}\,\bf ls,
\label{26}
\end{equation}
\begin{multline}
F^{(\ta)}=\frac{1}{r^2}\frac{\pt f_\text{ls}(r,\ta,\ffi)}{\pt\ta}\Biggl[
s_x\Biggl(\cos\ffi\,l_z-i\sin\ffi\,r\frac{\pt}{\pt r}\Biggr)+\\
s_y\Biggl(\sin\ffi\,l_z +i\cos\ffi\,r\frac{\pt}{\pt
r}\Biggr)+\,\cot\ta\,s_zl_z\Biggr],
\label{27}
\end{multline}
\begin{multline}
F^{(\ffi)}=\frac{1}{r^2}\frac{\pt f_\text{ls}(r,\ta,\ffi)}{\pt\ffi} \times \\
\Biggl[
s_x\Biggl(\cos\ffi(\sin\ffi\,l_x-\cos\ffi\,l_y)
-i\cot\ta\cos\ffi\,r\frac{\pt}{\pt r}\Biggr)+\\
s_y\Biggl(\sin
\ffi(sin\ffi\,l_x-\cos\ffi\,l_y)
-i\cot\ta\sin\ffi\,r\frac{\pt}{\pt r}\Biggr)+\\
s_z\Biggl(\cot\ta(sin\ffi\,l_x-\cos\ffi\,l_y)+i\,r\frac{\pt}{\pt
r}\Biggr)\Biggr].
\label{28}
\end{multline}
It can be seen from \eqref{25}--\eqref{28} that for a spherical nucleus, when $f_\text{ls}$ does not depend on $\ta$ and $\ffi$, the spin-orbit potential \eqref{21} reduces to its usual form, given by the second term in Eq. \eqref{1}.

With the help of the expansion
\begin{equation}
\frac{1}{\sqrt{r^2+(r')^2-2rr'\cos\beta}}=\sum_{\lm=0}^\infty
k_\lm(r,r')P_\lm(\cos\bt),
\label{30}
\end{equation}
where
\begin{equation}
k_\lm(r,r') =
\begin{cases}
r^\lm/(r')^{\lm+1} &
\text{for $r\le r'$}, \\
(r')^\lm/r^{\lm+1} & \text{for $r>r'$},
\end{cases}
\label{31}
\end{equation}
which follows from the generating function of Legendre polynomials, expression \eqref{29} for the Coulomb potential can be transformed to the form 
\begin{multline}
V_\text{Coul}(r,\ta,\ffi)=
\frac{3qZe^2}{R_\text{Coul}^3}\sum_{\lm=0}^\infty
\sum_{\mu=-\lm}^\lm(2\lm+1)^{-1} Y_{\lm\mu}(\ta,\ffi)\,\times\\
\int\limits_0^{2\pi}d\ffi'\int\limits_0^{\pi}
Y^*_{\lm\mu}(\ta',\ffi')K_\lm(r,\ta',\ffi')\sin\ta'd\ta',
\label{32}
\end{multline}
where
\begin{multline}
\label{34}
K_\lm(r,\ta,\ffi)= \\
\begin{cases}
\frac{(2\lm+1)r^2}{(\lm+3)(\lm-2)}
-\frac{r^\lm(\lm-2)^{-1}}{R_\text{Coul}^{\lm-2}(\ta,\ffi)}&\text{if $R_\text{Coul}(\ta,\ffi)\ge r$, $\lm\ne2$};\\
\frac{r^2}{5}+r^2\ln\left(\frac{R_\text{Coul}(\ta,\ffi)}{r}\right)&\text{if $R_\text{Coul}(\ta,\ffi)\ge r$, $\lm=2$};\\
\frac{1}{\lm+3}\frac{R_\text{Coul}^{\lm+3}(\ta,\ffi)}{r^{(\lm+1)}}
&\text{if $R_\text{Coul}(\ta,\ffi)<r$.}
\end{cases}
\end{multline}

It can be checked that the function \eqref{34} is continous at $R_\text{Coul}(\ta,\ffi)=r$. When $\bt=\gm=0$ the obtained relationships \eqref{32}, \eqref{34} reduce themselves to the usual expression \eqref{5} for the Coulomb potential of a uniformly charged sphere.

\section{Computation of single-particle states}

Eigenstates of anisotropic single-particle Hamiltonian
\begin{equation}
H=T+V(r,\ta,\ffi),
\label{35}
\end{equation}
where $T=-\frac{\hbar^2}{2M}\Delta$ is the kinetic energy operator, are calculated by diagonalization of the Hamiltonian matrix in the basis of isotropic harmonic oscillator
\begin{equation}
\la{\bf r}\sigma|N lms\ra=U_{N l}(r)Y_{lm}(\ta,\ffi)
\la\sigma|s\ra,
\label{36}
\end{equation}
where $N$ is the number of excited quanta, $l$ is the orbital angular momentum of nucleon, $m$ and $s$ are the projections of, respectively, the orbital angular momentum and spin of nucleon along the $z$ axis, $U_{Nl}(r)$ is the radial part of harmonic oscillator wavefunction, $Y_{lm}(\ta,\ffi)$ is the spherical harmonic, and $\la\sigma|s\ra$ is the spin part of the wavefunction.

The usual expression from Ref. \cite{nilsson-1955} for the energy of the oscillator quantum $\om=41A^{-1/3}$ was used. During diagonalization of the Hamiltonian matrix it was possible to considerably reduce computation time due to conservation of parity $\pi$ which follows from the spatial inversion symmetry of ellipsoidal potential. Calculations were performed in truncated oscillator basis with $N_\text{max}=11$. This choice of truncation was checked to be reliable enough for description of bound states of neutrons and protons in ellipsoidal non-axial potential in the range of masses $0 \le A \lsim 240$ with deformations $-0.5 \le \beta \le 0.5$. Further increase of $N_\text{max}$ had only negligibly small effect on the results.


By usage of the properties of spherical harmonics it is easy to show that calculation of the matrix elements $\la N'l'm's'|V_i|Nlms\ra$ is reduced to computation of integrals of the form
\begin{multline}
I_1=\int\limits_0^\infty U_{N'l'}(r)U_{Nl}(r)r^2dr \\
\int\limits_0^{2\pi}d\ffi\,
\int\limits_0^\pi G_1(r,\ta,\ffi)
 Y^*_{\lm\mu}(\ta,\ffi)\sin\ta d\ta\nonumber
\end{multline}
\begin{multline} 
I_2=\int\limits_0^\infty U_{N'l'}(r)\frac{\pt U_{Nl}(r)}{\pt
r}r^2dr \\
\int\limits_0^{2\pi}d\ffi\,
\int\limits_0^\pi G_2(r,\ta,\ffi)
 Y^*_{\lm\mu}(\ta,\ffi)\sin\ta d\ta
 \label{37}
\end{multline}
where the functions $ G_1(r,\ta,\ffi)$,  $ G_2(r,\ta,\ffi)$, are connected by simple relations with functions $f_\text{nucl}(r,\ta,\ffi)$,
$\frac{1}{r}\frac{\pt f_\text{ls}(r,\ta.\ffi)}{\pt r}$, $\frac{\pt f_\text{ls}(r,\ta,\ffi)}{\pt\ta}$, $\frac{\pt f_\text{ls}(r,\ta,\ffi)}{\pt\ffi}$ and $K_\lm(r,\ta,\ffi)$; $\lm$ takes even values in the interval $[|l'-l|,l'+l]$ and $\mu=m'-m,m'-m\pm 1$.

For evaluation of the triple integrals $\eqref{37}$ the following method was used. First, on a sufficiently fine grid of the values $r_i$ with the step of 0.1 fm the double integrals over angular variables, which determine the coefficients $\al_{\lm\mu}(r_i)$ of expansion of the functions $G_{1,2}(r,\ta,\ffi)$ into spherical harmonics at $r_i$, were evaluated using the SHTNS library \cite{shtns}, after which the expansion coefficients were used for numerical computation of radial integrals on the grid $r_i$.

In computation of $\al_{\lm\mu}(r)$ one has to calculate only the values at $\mu \ge 0$, and then the symmetry property can be used:
\begin{equation}
\al_{\lm,-\mu}(r)=(-1)^\mu\al^*_{\lm,\mu}(r).
\label{41}
\end{equation}

\section{Parameterization of surface diffuseness}

\label{section3}

For completion of the description of the deformed one-particle potential \eqref{6} it is necessary to parameterize the form factors $a'_i(\bt,\gm)$ ($i=\text{nucl, ls}$) in the expression of the surface diffuseness \eqref{13}. We begin with demonstration that the effect of variation of the diffuseness of the nuclear surface at different deformations in fact takes place. Earlier, in Introduction it has been mentioned that the smallest value of the surface diffuseness should be expected near the stable equilibrium deformation, where an isolated dense group of single-particle levels is formed near the Fermi surface: at $\beta=0$ for nuclei with closed outer shells and at certain $\beta_\text{eq} \ne 0$ in deformed nuclei. Consequently, deformation of spherical nuclei with completely or almost filled outer shells should result in increase of the surface diffuseness, while in deformed nuclei as one approaches to $\beta \to \beta_\text{eq}$ a decrease of the surface diffuseness should be observed.

We will illustrate this on the example of a spherical nucleus $^{52}$Cr with a magic neutron number $N=28$ and a deformed spheroidal nucleus $^{76}$Sr, having a singularity in the distribution of single-particle levels near the Fermi surface at $\beta \sim$ 0.3--0.4. For this purpose we calculate for each of these nuclei the gradient of the nucleon density $\nbl\rho(\bf r)$ at the surface at different values of the deformation parameter $\beta$. The value of the gradient serves as an indicator of the value of the diffuseness: smaller absolute values of the gradient correspond to larger diffuseness, and vice versa (see, e.g., Eq. \eqref{12}). At the nuclear surface the gradient of the nucleon density has the largest negative value. Due to the requirement that this value is constant at all points of the surface, for the considered axially-symmetric nuclei it can be found as the minimum of the derivative $\frac{\pt\rho(r,\ta=0,\ffi=0)}{\pt r}$ taken along the symmetry axis, perpendicular to the nuclear surface. The corresponding derivative is determined by the expression
\begin{multline}
\frac{\pt\rho(r,0,0)}{\pt r}= \\
\sum_{i=1}^A\sum_{Nlms}\sum_{N'l'm'} c_i(Nlms)c_i(N'l'm's) \, \times \\
\frac{\pt [U_{Nl}(r)U_{N'l'}(r)]}{\pt r} 
Y_{lm}(0,0)Y_{l'm'}(0,0),
\label{a}
\end{multline}
where $c_i(Nlms)$ are the coefficients of expansion of the eigenstates of the Hamiltonian \eqref{35}, occupied by nucleons in the ground state of the nucleus, into basis wavefunctions \eqref{36}.

\begin{figure}[tb]
    \centering
    \includegraphics[width=8cm]{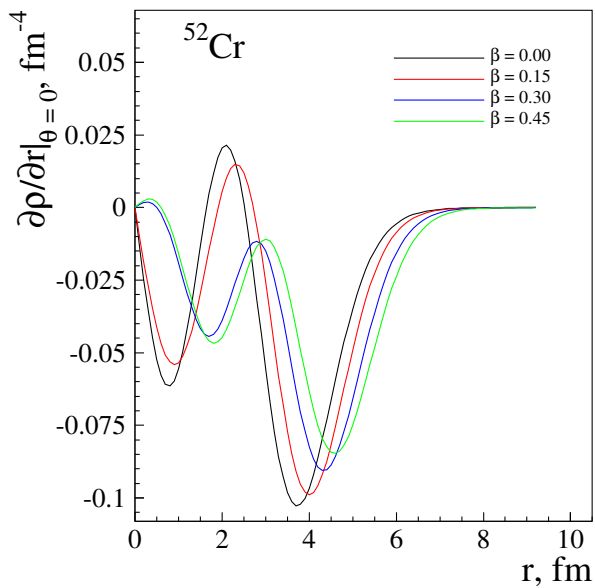}
    \caption{Dependence of the radial derivative of nucleon density along the nuclear symmetry axis on the deformation parameter $\beta$ for $^{52}$Cr. The minumum of the derivative observed at $r=R_\text{nucl}(\ta=0,\ffi=0)$ is equal to the gradient of the nucleon density at the surface.}
    \label{fig1}
\end{figure}

\begin{figure}[tb]
    \centering
    \includegraphics[width=8cm]{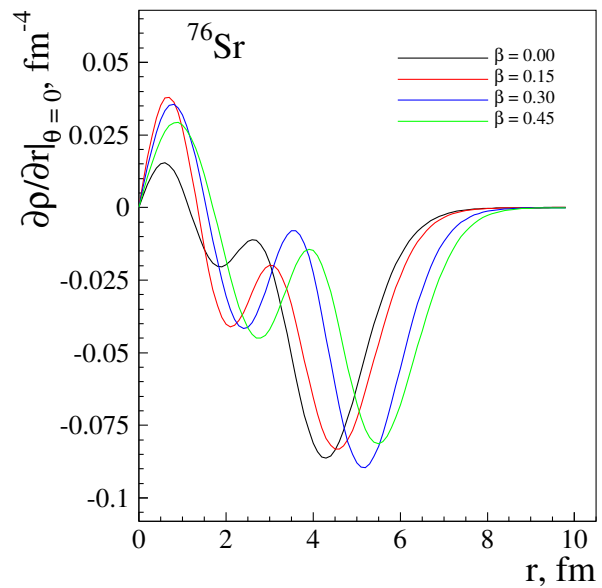}
    \caption{Dependence of the radial derivative of nucleon density along the nuclear symmetry axis on the deformation parameter $\beta$ for $^{76}$Sr. The minumum of the derivative observed at $r=R_\text{nucl}(\ta=0,\ffi=0)$ is equal to the gradient of the nucleon density at the surface.}
    \label{fig2}
\end{figure}

Figures \ref{fig1}, \ref{fig2} show the derivatives $\frac{\pt\rho(r,\ta=0,\ffi=0)}{\pt r}$ calculated for $^{52}$Cr and $^{76}$Sr at different values of the deformation parameter $\beta$ (with the same value of the diffuseness parameter of the Woods--Saxon potentials at different $\beta$, i.e., $a_i'(\beta,\gamma) = a_i$ in Eq. \eqref{13}). It is seen from the figures, that for the spherical semi-magic $^{52}$Cr deformation results, as expected, in decrease (by absolute value) of the gradient of the nucleon density at the surface and, consequently, in increase of the diffuseness. At the same time, for the deformed $^{76}$Sr the minimal surface diffuseness correponds to the region of the equilibrium deformation (at $\beta_2 \approx 0.3 \div 0.4$ \cite{nacher-2004,cdfe}), and it increases both to the left and to the right of this region. Thus, both for spherical and deformed nuclei in the region of the equilibrium deformation decreased diffuseness of the nuclear surface, indeed, takes place, which results in decrease of the surface energy and supports the stability of equilibrium shape. It follows that it is natural to vary the diffuseness parameter of the realistic potential to adjust the behaviour of the deformation energy.

In what manner should the diffuseness parameter be varied? Clearly, determination of the true dependence of the diffuseness parameter on the nuclear deformation parameters is hardly possible, since it would effectively mean an attempt of phenomenological construction of the self-consistent mean field. Nevertheless, description of the deformation energy can be significantly improved by a reasonable strategy of adjustment of the diffuseness parameter in the process of deformation. Thus, for nuclei with nearly closed shells as deformation increases it is beneficial to use a slightly increasing diffuseness parameter, which will result in additional stabilization of the spherical shape. On the other hand, for nuclei with partially filled outer shells having a stable deformation one should use a smoothly decreasing diffuseness parameter, which, in the region of formation of the equilibrium deformation, will result in an additional vertical shift of the local minimum in the right direction. The amount of the variation can be adjusted from the experimental data, which are rather reliable for such nuclei. A smoothly varying diffuseness curve can not produce a false minimum of the potential energy by itself, and, therefore, the correction, introduced for deformed nuclei, can be relatively safely used also for transitional nuclei, having as a rule a significant dynamic deformation.

We assume that variation  a diffuseness parameter $a'_i(\bt,\gm)$ from its value  $a_i$ in a spherical nucleus is of the order $O(\beta)$ and express it in the form
\begin{equation}
a'_i(\bt,\gm)=a_i(1+k\bt\psi(\gm')),\hspace{0.5cm}i=\text{nucl},\text{ls},
\label{14}
\end{equation}
where $k$ is a parameter governing the magnitude and sign of the considered effect, $\gm'=\gm/(\pi/3)$ is the normalized value of the non-axiality parameter taking values in the range $[0,1]$, and $\psi(\gm') \sim 1$ is a positive-valued function describing the dependence of the diffuseness parameter on the non-axiality.

As already mentioned, we assume that $k > 0$ if the number of neutrons or protons in nucleus is sufficiently close to the magic number. 
In this case the function $\psi(\gm')$ is unimportant and can be put equal to 1. For nuclei with partially filled shells (deformed and transitional nuclei) we assume that $k < 0$ and expand  the function $\psi(\gm')$ into series up to the 3-rd order:
\begin{equation}
\psi(\gm')=1+p_1\gm'+p_2\gm'^2+p_3\gm'^3.
\label{15}
\end{equation}

It is seen from \eqref{14} and \eqref{15} that for prolate ($\gamma = 0$) and oblate ($\gamma = \pi/3$) spheroidal deformations the factor $ a'_i$ takes the form
\begin{equation}
a'_i(\bt,\gm) = 
\begin{cases}
a_i(1+k\bt) & \hspace{0.5em}\text{for $\gm=0$}, \\
a_i(1+kq\bt) & \hspace{0.5em}\text{for $\gm=\pi/3$},
\end{cases}
\label{16}
\end{equation}
where
\begin{equation}
q=1+p_1+p_2+p_3>0.
\label{17}
\end{equation}
Thus, the relative reduction of the diffuseness parameter, leading to increased tendency of deformation, for prolate nuclei is determined by the value $k\beta$ and for oblate $kq\beta$. It can be expected that $q < 1$, since most deformed nuclei are prolate.

It is known from experiment that the majority of nuclei in ground state are axially-symmetric. It follows from \eqref{14} and \eqref{15} that in order to reproduce this behavior the derivative
\begin{equation}
\frac{\pt\psi(\gm')}{\pt\gm'}= p_1+2p_2\gm'+3p_3\gm'^2
\label{18}
\end{equation}
should be negative at $\gm'=0$ (which implies that $p_1 < 0$) and positive at $\gm'=1$. This choice of the parameter values inhibits motion of the minimum of the deformation energy away from the $\gamma = 0$ or $\gamma = \pi/3$ line, corresponding to axial symmetry.

Assuming that the $\psi(\gm')$ function acts similarly on prolate and oblate spherical shapes, so that $\frac{\pt\psi(0)}{\pt\gm'}
=-\frac{\pt\psi(1)}{\pt\gm'}$ we obtain the equation for the $p_{1,2,3}$ parameters:
\begin{equation}
2p_1+2p_2+3p_3=0.
\label{19}
\end{equation}
And from \eqref{17} and \eqref{19}
\begin{align}
\label{20}
p_2 &= -3(1-q)-p_1\\
p_3 &= 2(1-q). \nonumber
\end{align}

Therefore, the parameter set used for adjustment of the effect of surface energy is comprised of the values $k$, $q$, $p_1$.

\section{Effect of pairing interaction}

As usual, pairing is treated separately for neutrons and protons, which is justified by the notion that neutron-proton correlations of superconductive kind can be neglected in complex nuclei \cite{soloviev-1963}. Since the obtained equations do not differ for protons and neutrons, in the following we do not specify explicitly the type of nucleons.

The Hamiltonian \eqref{35}, corresponding to an ellipsoidal nucleus, is invariant under time reversal. It implies that its single-particle levels $e_k$ are doubly degenerate. Each of the degenerate levels corresponds to a pair of conjugate states $|k\sigma\rangle$, $\sigma=+,-$. To take into account the pairing effect we use a version of the BCS model \cite{bardeen-1957,*bardeen-1957b,belyaev-1959,soloviev-1959,*soloviev-1959b} in which it is assumed that the residual pairing interaction takes place only between pairs of conjugate states and the interaction matrix element is constant for any transitions between states. In this case the Hamiltonian of a system of nucleons in the average nuclear field at the presence of pairing interaction has the form
\begin{eqnarray}
\HH=\sum_{k\sg}e_k a^+_{k\sg}a_{k\sg}-G\sum_{k\ne k'}a^+_{k+}a^+_{k-}
a_{k'-}a_{k'+},
\label{49}
\end{eqnarray}
where the diagonal terms of the pairing interaction are omitted, as they enter the average field, $G$ is the interaction constant and $a^+_{k\sg}$ and $a_{k\sg}$ are the operators of creation and destruction of a nucleon in the state $|k\sigma\rangle$ with energy $e_k$, satisfying the anticommutativity condition.

The matrix elements of the pairing interaction can be taken as approximately constant only in a limited region of single-particle energies $e_k$, and the indices $k$ and $k'$ can take values only in a limited range $[N_1, N_2]$, where the energies of the $N_1$-th and $N_2$-th single-particle levels are, respectively, less and greater than the Fermi energy $e_F$. Moreover, if the number of nucleons in the system is odd, the single-particle level at $e_F$ with $k=N_F$ must be excluded from \eqref{49}, since the nucleon at this level prevents a bound nucleon pair from occupying it (the blocking effect). Thus, the Hamiltonian \eqref{49} describes only the subsystem of $n$ nucleons occupying upper-lying filled levels in the neighborhood of the Fermi surface, where $n=N-2N_1+2$ for even $N$, and $n=N-2N_1+1$ for odd $N$.

The effect of the pairing interaction can be treated approximately if using the Bogoliubov transformation one moves from the particle creation and destruction operators to the corresponding operators of quasiparticles:
\begin{equation}
a_{k\sg}=u_k\al_{k-\sg}+\sg v_k\al^+_{k\sg},
\label{50}
\end{equation}
where the following condition is assumed to hold:
\begin{equation}
u^2_k+v^2_k=1.
\label{51}
\end{equation}

The transformation \eqref{50}, \eqref{51} breaks conservation of the number of particles in the considered nucleon subsystem. Therefore, the condition of conservation of the average number of particles is introduced:
\begin{equation}
\la\Psi_0|\hat n|\Psi_0\ra=2\sum_kv^2_k=n,
\label{53}
\end{equation}
where $|\Psi_0\ra$ is quasiparticle vacuum state and $\hat n=\sum_{k\sg}a^+_{k\sg}a_{k\sg}$ is the number of particles operator.

To find the system of equations of the BCS method one has to minimize the average energy of the quasiparticle vacuumu $\la\Psi_0|\HH|\Psi_0\ra$ via variation of the parameters $u_k$, $v_k$, $k = N_1,\dots,N_2$ with respect to the constraints \eqref{51}, \eqref{53}. With the help of the Lagrange multiplier method the following equations are obtained:
\begin{equation}
n =\sum_{k=N_1}^{N_2}\Biggl[1-\frac{\eps_k-\lm}
{\sqrt{(\eps_k-\lm)^2 +\DD^2}}\Biggr],
\label{63}
\end{equation}
\begin{equation}
\frac{2}{G}=\sum_{k=N_1}^{N_2}\frac{1}{\sqrt{(\eps_k-\lm)^2 +\DD^2}}.
\label{64}
\end{equation}
\begin{equation}
v^2_k=\frac{1}{2}\Biggl[1-\frac{\eps_k-\lm}{\sqrt{(\eps_k-\lm)^2
+\DD^2}}\Biggr], \hspace{0.5cm}k=N_1,N_1+1,\ldots,N_2,
\label{60}
\end{equation}
\begin{equation}
\eps_k= e_k-Gv^2_k, \hspace{0.5cm}k=N_1,N_1+1,\ldots,N_2,
\label{59}
\end{equation}
where $\eps_k$ are the renormalized single-particle energies, $\lambda$ is the chemical potential which determines the energy of the uppermost occupied level,
\begin{equation}
\DD=G\sum_ku_kv_k
\label{57}
\end{equation}
is correlation function commonly called the pairing gap. It can be shown (see, e.g., \cite{soloviev-1959}) that the excited states of a system containing an even number of nucleons are separated from the ground state by an energy interval approximately equal to $2\Delta$. Pairing gap $\DD$ can be estimated directly from experimental nuclear masses (in contrast with the interaction constant $G$).

Equations \eqref{63}--\eqref{59} form a system of equations, by solving which one obtains the values $v_k$ and also $\lambda$ and $\Delta$ (if the constant of interaction $G$ is known) or $\lambda$ and $G$ (if, instead, the pairing gap $\Delta$ is specified).

The correction, that has to be introduced to the total single-particle energy $E$ of all nucleons of same kind to incorporate the pairing effect, has the form
\begin{multline}
\DD E= \\
\sum_{k=N_1}^{N_2}(2v^2_k-n_k)e_k-\frac{\DD^2}{G}
-G\sum_{k=N_1}^{N_2}v^4_k+\frac{1}{2}G\sum_{k=N_1}^{N_2}n_k,
\label{66}
\end{multline}
where $n_k$ is the number of nucleons occupying the given doubly-degenerate level without pairing, taking the values 2 or 0, if, respectively $2k \le N_F$ and $ 2k > N_F$, and, in order to take into account the blocking effect, $k\ne N_F$ for odd $N$.

\section{Choice of the model parameters and its application to calculating the deformation of medium and heavy nuclei}

At the final step of construction of the model we describe the choice of the numerical parameters of the dependence of the diffuseness of the surface on deformation. In the present work we perform an approximate estimation of their values based on a limited subset of 36 medium and heavy nuclei. The nuclei in the subset (see table \ref{table1}) were chosen on the basis of availability of information in the compilation of static electric quadrupole moments \cite{stone-2005} and database of experimental deformation parameters \cite{cdfe} derived from it, so that to include most important types of ellipsoidal deformations at $A \ge 50$. In future the parameters of the model will be refined by performing calculations for a larger array of isotopes and by using one of the statistical optimization methods.

For each isotope the deformation was estimated by calculating the potential energy of deformation $E$ on a grid in the range $0 \le \beta \le 0.5, 0 \le \gamma \le \pi/3$. For this, the $(\beta,\gamma)$ plane was uniformly divided into $30 \times 30$ intervals, and the equilibrium ground-state deformation parameters were determined as the location of the absolute minimum of $E$.

During calculation of the pairing correction we used all (doubly degenerate) single-particle energy levels from $N_1=1$ to $N_2=2N_F$. This choice is helpful for unification of the calculation, while at the same time the nucleons occupying deep-lying levels mainly result in renormalization of the interaction constant $G$ \cite{soloviev-1959} without significant alteration of the effect of the pairing correction on the calculated deformations. The pairing gap $\Delta$ was estimated from experimental masses of neighboring using the difference formulae from \cite{moeller-1992}. Thus, for each deformation $(\beta,\gamma)$ the values of the interaction constant $G$, chemical potential $\lambda$, and $v_k$, $u_k$ ($k=N_1,N_1+1\dots,N_2$) were determined from the pairing equations.

A commonly used formulation of the nuclear shape is in terms of the multipole deformation parameters $\beta_\lambda$. For comparison of our results with the available data, for the equilibrium values of $\beta$ and $\gamma$ we calculated the corresponding values of $\beta_2$ and $\beta_4$ (see Appendix). As seen from tables 1 and 2, in the considered region of deformations the absolute value of $\beta_2$ is approximately equal to~$\beta$.

As explained in Section III, determination of the model parameters was performed separately for two groups of isotopes: nuclei with $N=N_\text{mag},N_\text{mag} \pm 1$ or $Z=Z_\text{mag},Z_\text{mag} \pm 1$ (where $N_\text{mag},Z_\text{mag}$ are magic numbers for spherical nucleus), and the rest of nuclei with partially filled shells.

(a) First, we consider deformed and transitional nuclei, having partially filled outer-most shells. For such nuclei the parameters $k$, $q$, and $p_1$ have to be estimated.

\begin{figure}[htb]
    \centering
    \includegraphics[width=\columnwidth]{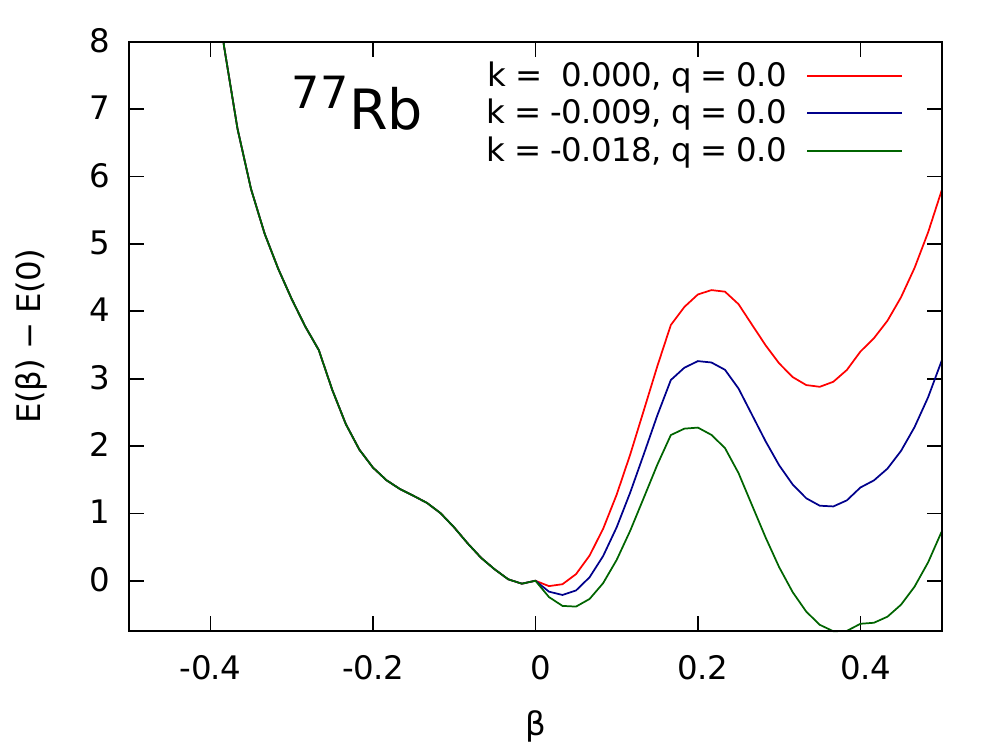}
    \caption{Effect of choice of the parameter $k$ on the potential energy of deformation of $^{77}$Rb in ground state.}
    \label{fig3}
\end{figure}

\begin{figure}[htb]
    \centering
    \includegraphics[width=\columnwidth]{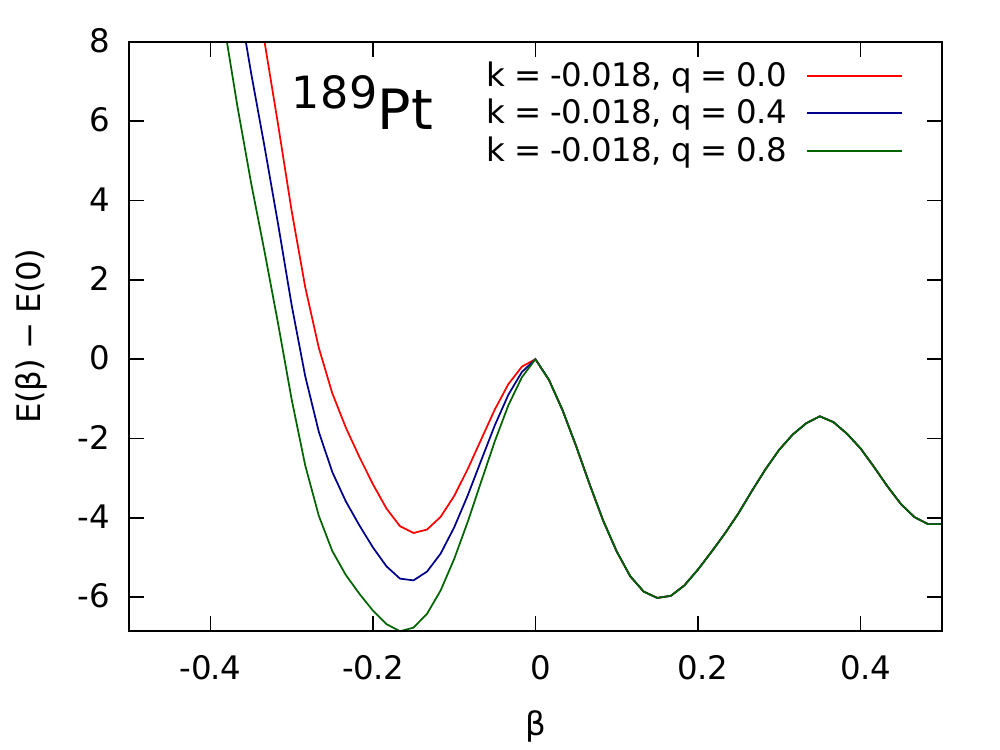}
    \caption{Effect of choice of the parameter $q$ on the potential energy of deformation of $^{189}$Pt in ground state.}
    \label{fig4}
\end{figure}

As seen from Eq. \eqref{16}, estimation of the parameters $k$ and $q$ can be performed using known data on nuclei with pronounced prolate and oblate spheroidal deformation. For prolate spheroidal nuclei the diffuseness parameter $a'_i(\bt,\gm)$, $i=\text{nucl},\text{ls}$, depends only on $k$. We gradually decreased the value of $k<0$ until it was possible to reproduce the deformation of not only rare-earth and actinide nuclei but also medium prolate spheroidal nuclei such as the isotopes of rubidium $^{76,77}$Rb, for which the experimental value of the deformation parameter $\beta_2 \approx 0.4$ could not be reproduced without the variation of diffuseness. Further decrease of $k$ is limited by competition between different local minima of the potential energy, corresponding to both positive and negative deformation, in such nuclei as the oblate $^{189}$Pt and prolate $^{181}$Ta, and the decrease of $k$ has to be compensated by increasing $q$ for correct description of oblate shapes. The parameter $q$ is restricted by the inequality $0<q<1$ (see Section \ref{section3}). Therefore, it was found that the parameters $k$ and $q$ can be varied only in a narrow interval of values: $-0.020 < k < -0.015$ and $0.5 < q < 1$. After this a least-squares search within the specified limits, including remaining nuclei from table 1, yielded the following optimal values of $k$ and $q$:
\begin{equation}
k=-0.018, \qquad q=0.8.
\label{67}
\end{equation}
Further variation of the obtained values of the parameters $k$ and $q$ by up to, respectively, 10\% and 5\%, has a little effect on the calculated deformations of the considered nuclei.

The parameter $p_1$ describes the relative preference of spheroidal (i.e. $\gamma = 0$ or $\gamma=\pi/3$) shapes. We chose it equal to
\begin{equation}
p_1 = -0.4.
\label{68}
\end{equation}
With this choice the majority of deformed nuclei are spheroidal in ground state, while on the other hand non-axial deformations still can occur when such tendency is present in the one-particle potential prior to variation of diffuseness.

(b) For the nuclei with $Z=Z_\text{mag},Z_\text{mag} \pm 1$ or $N=N_\text{mag},N_\text{mag} \pm 1$ we found the value
\begin{equation}
    k=+0.006,
    \label{69}
\end{equation}
which yields satisfactory agreement with the considered data. (The $q$ and $p_1$ parameters are not used, since the function $\psi(\gamma')$ is assumed to be equal to 1 for such nuclei.)

\begin{figure}[htb]
    \centering
    \includegraphics[width=\columnwidth]{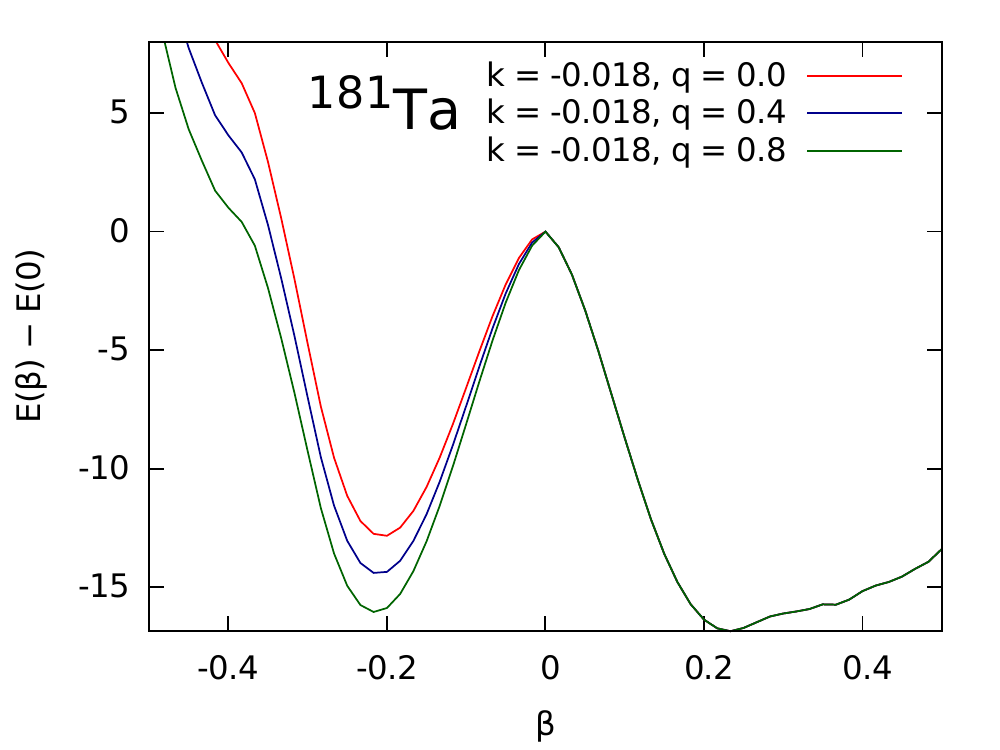}
    \caption{Effect of choice of the parameter $q$ on the potential energy of deformation of $^{181}$Ta in ground state.}
    \label{fig5}
\end{figure}

\begin{figure}[htb]
    \centering
    \includegraphics[width=\columnwidth]{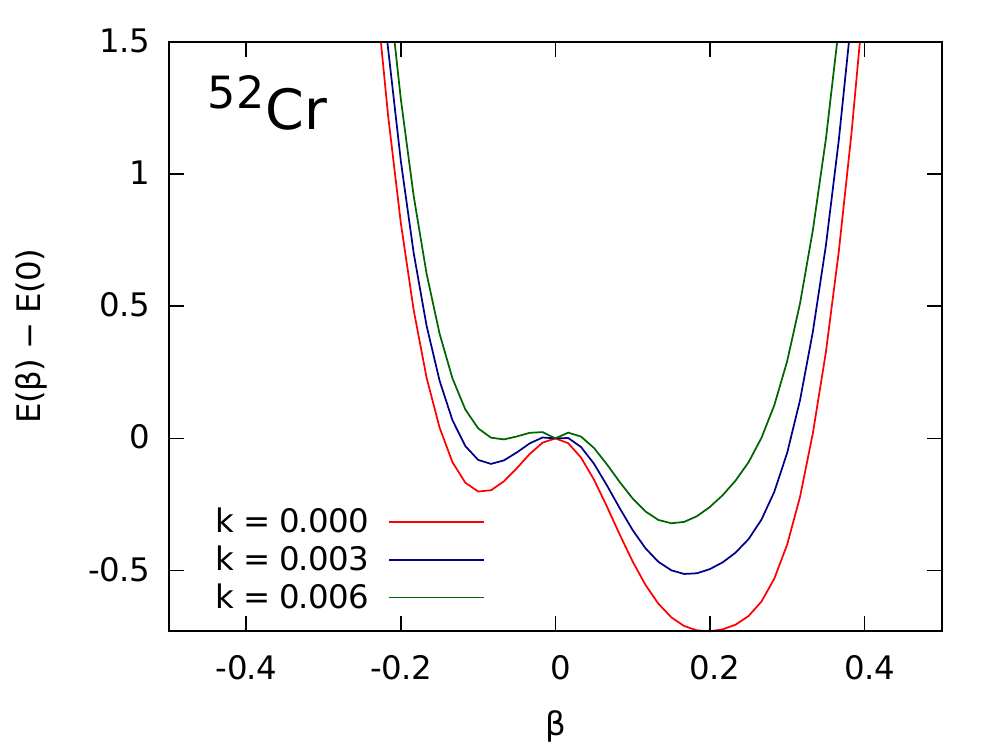}
    \caption{Effect of choice of the parameter $k$ on the potential energy of deformation of $^{52}$Cr in ground state.}
    \label{fig6}
\end{figure}

\begin{figure*}[htb]
    \centering
    \includegraphics[width=0.7\columnwidth]{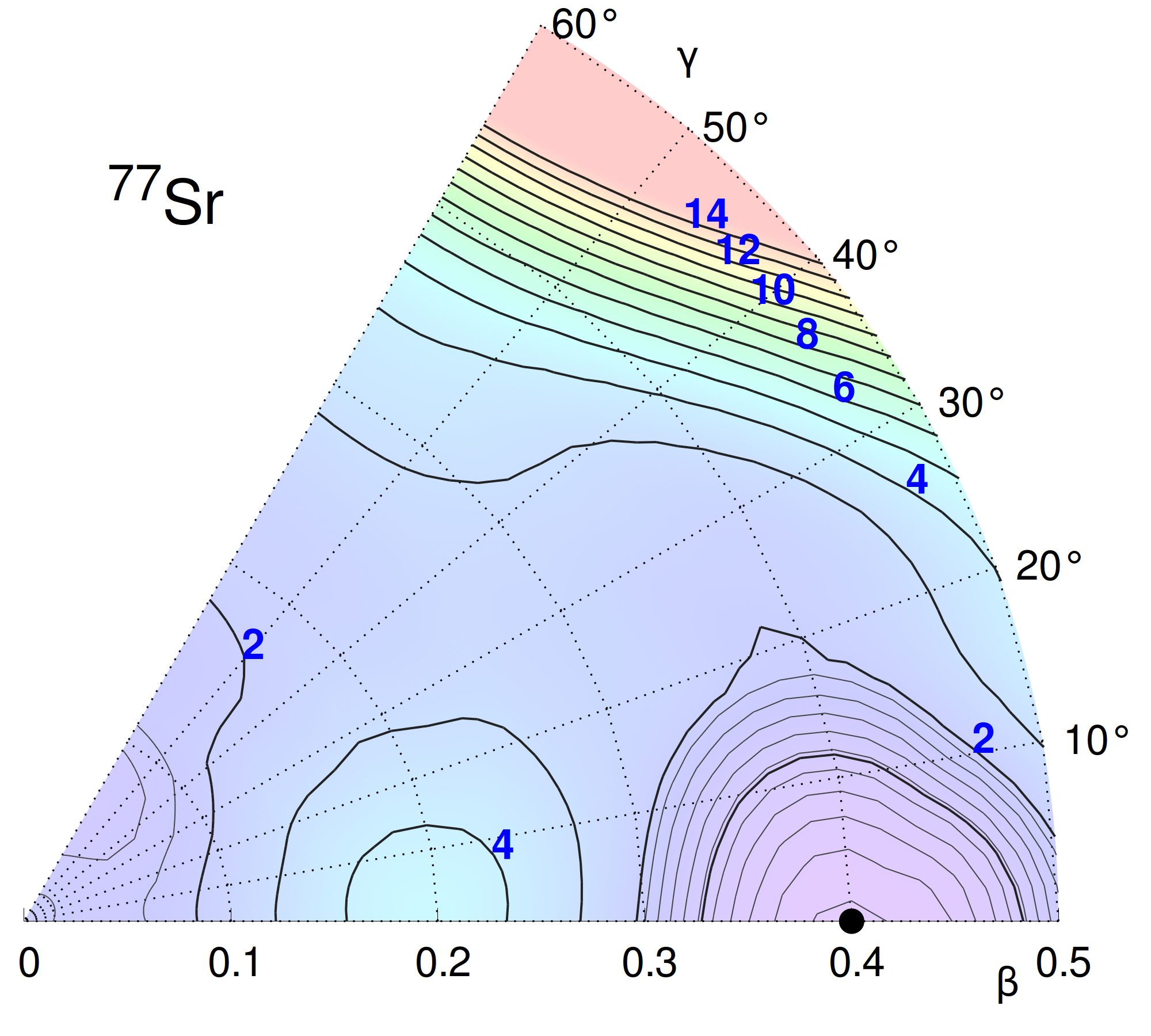}
    \includegraphics[width=0.7\columnwidth]{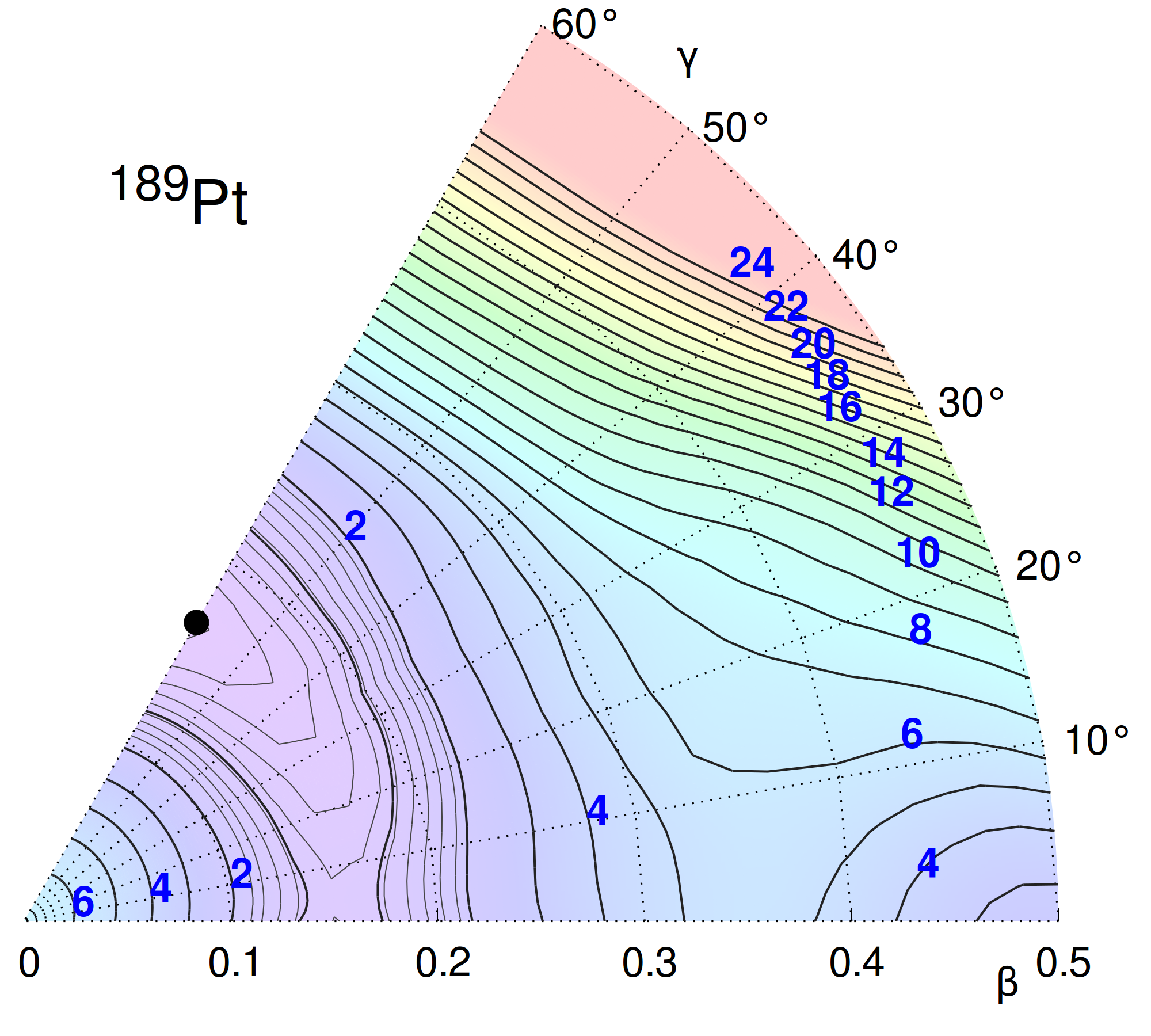}

    \includegraphics[width=0.7\columnwidth]{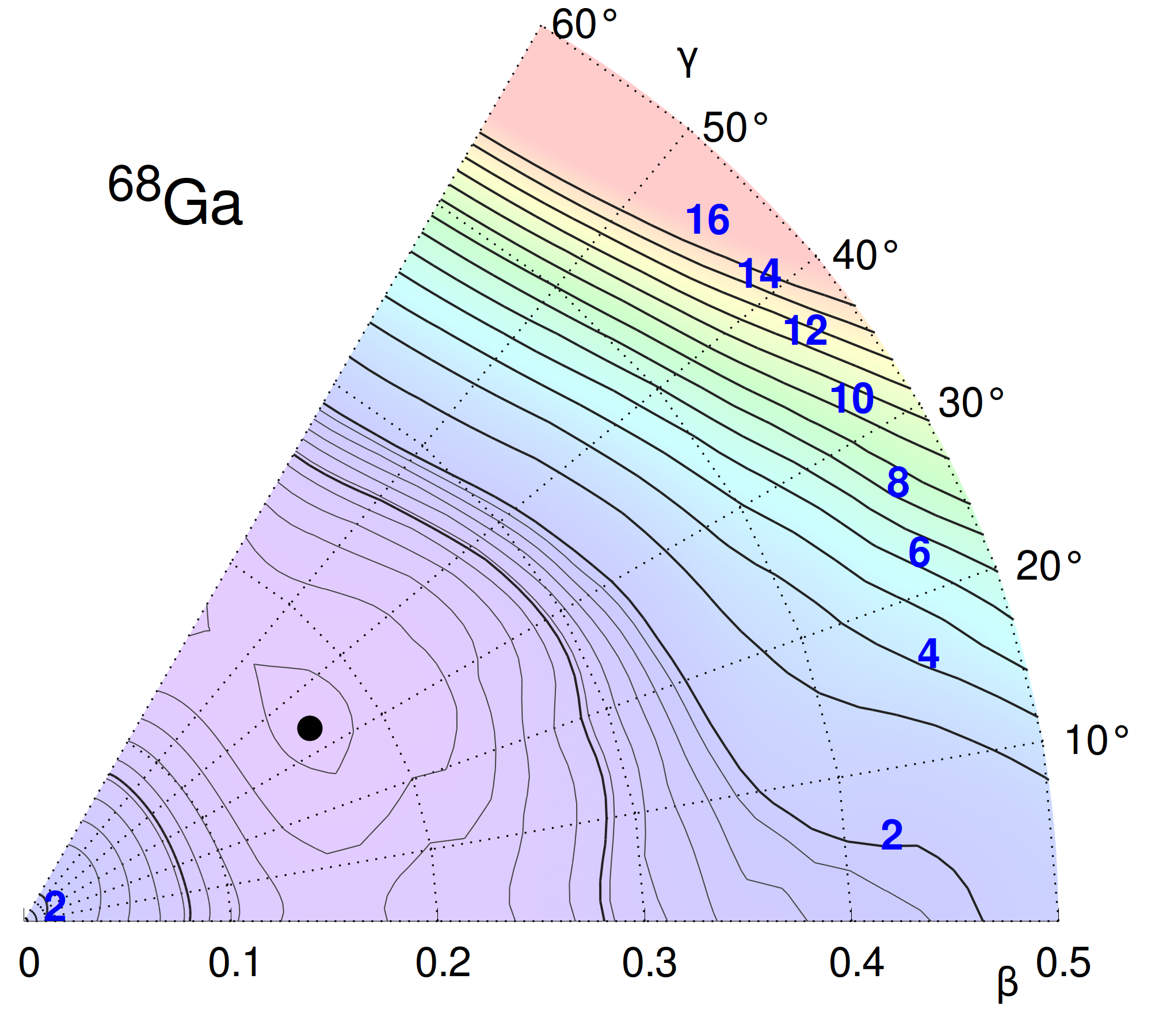}
    \includegraphics[width=0.7\columnwidth]{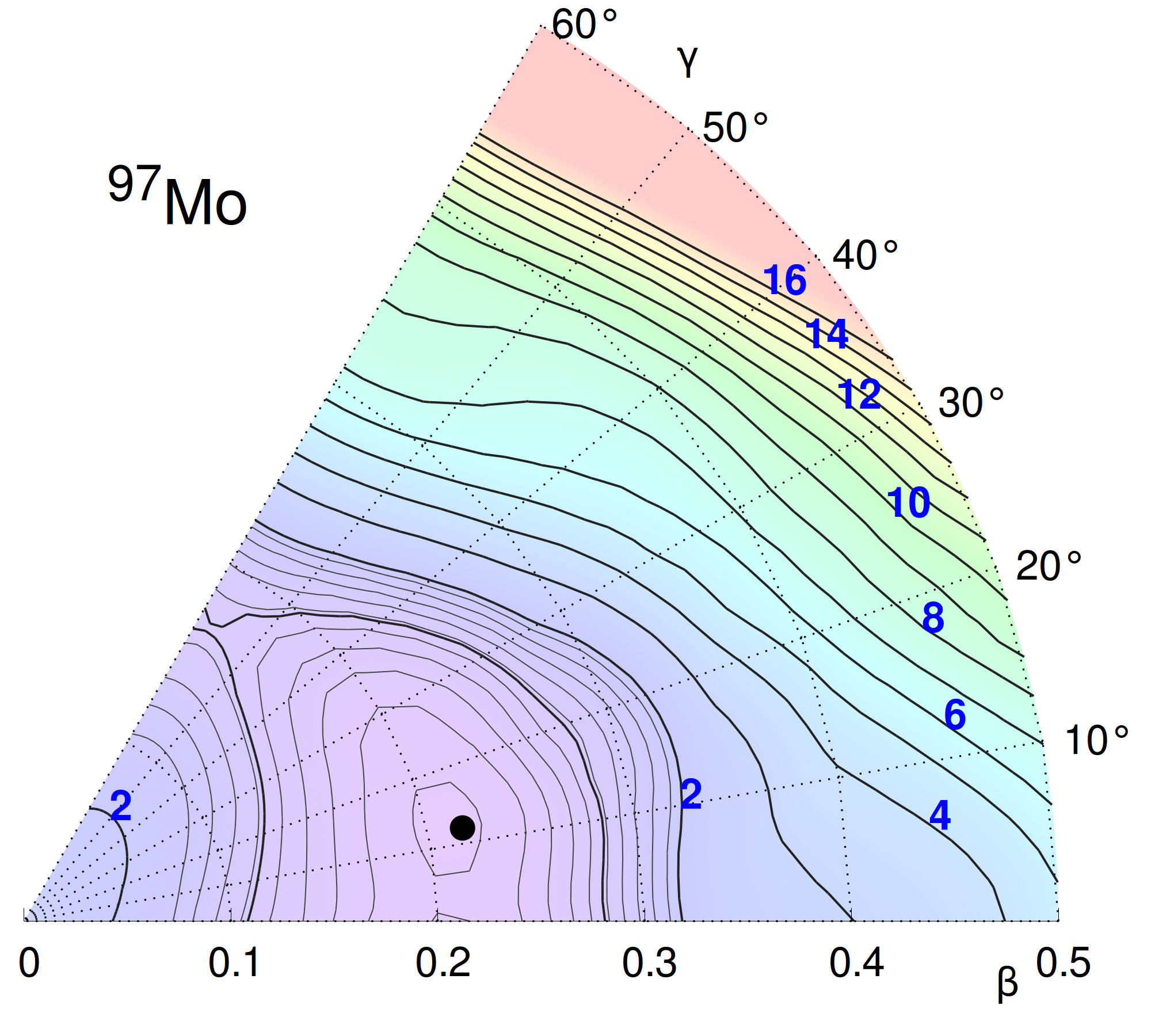}
    \caption{Potential energy of deformation of $^{77}$Sr, $^{198}$Pt, $^{68}$Ga, $^{97}$Mo in the $(\beta,\gamma)$ plane. Thick contour lines are drawn in 1 MeV steps, thin lines---in 0.2 MeV steps. Energies in MeV, relative to the absolute minimum value, are shown near thick contour lines. Black circle denotes location of the absolute minimum.}
    \label{fig7}
\end{figure*}

Overall, it should be emphasized that in the considered interval of deformations $-0.5 \le \beta \le 0.5$ the specified values of the parameters $k$, $q$, and $p_1$ lead to less than 1\% variation of the diffuseness. However, this variation has a very substantial effect on the calculated ground-state shape of nuclei. This is illustrated in Figs. \ref{fig3}--\ref{fig6} with the help of one-dimensional plots of the dependence of the potential energy of deformation on the value of $\beta$ in the spheroidal case, corresponding in the $(\beta,\gamma)$ plane to evolution of $E$ along the $\gamma=0$ line ($\beta > 0$ region in the figures) and along the $\gamma=\frac{\pi}{3}$ line ($\beta < 0$ region).

Thus, Fig. \ref{fig3} demonstrates behavior of the potential energy of deformation calculated for the spheroidal prolate nucleus $^{77}$Rb for different $k$. It is seen that without variation of diffuseness ($k=0$) the absolute minimum is formed in the potential energy curve, at $\beta \approx 0$ leading to incorrect prediction of almost spherical shape of this strongly deformed isotope (see table \ref{table1}). With variation of diffuseness, as the parameter $k$ changes from 0 to $-0.018$ the depth of the local minimum at $\beta \approx 0.4$ increases and begins to determine the resulting ground-state equilibrium deformation, close to the experimental value $+0.4$ \cite{harder-1996,harder-1997}.

In Figs. \ref{fig4} and \ref{fig5} the situation of competition of two local minima, only one of which corresponds to the experimentally expected ground-state deformation in spheroidal $^{189}$Pt and $^{181}$Ta, is shown. In this case one of the minima is at $\beta > 0$, the other one being at $\beta < 0$. With $q=0$ both isotopes have prolate deformation in ground state, while $^{189}$Pt is expected to be oblate from the value of the quadrupole moment \cite{stone-2005}. It is seen that the disagreement is resolved with the help of the parameter $q$. It can be noted that its action only affects the negative deformation region of the potential energy curve, leaving intact the $\beta > 0$ region.

Figure \ref{fig6} shows the effect of variation of the diffuseness parameter on the potential energy of $^{52}$Cr with a magic neutron shell $N=28$. In \cite{koning-2003} this nearly spherical nucleus was included in the list of isotopes on the basis of which the global optical potential was constructed. However, as seen from Fig. \ref{fig6}, without variation of diffuseness the predicted value of g.s. deformation is approximately $\beta \approx +0.2$, corresponding to a pronounced prolate spheroidal shape. Without pairing a deep minimum at $\beta \approx +0.3$ is formed by the proton component of the total internal energy $E$, while the neutron part contributes a broad minimum at $\beta=0$, so reduction of the predicted deformation is a combined result of both consideration of pairing and variation of diffuseness.

Figure \ref{fig7} shows calculated two-dimensional surfaces of the deformation energy in the $(\beta,\gamma)$ plane for nuclei with different ellipsoidal shapes in the ground state: prolate axial ($^{77}$Sr), oblate axial ($^{198}$Pt), and non-axial ($^{68}$Ga, $^{97}$Mo). In the figures, thick contour lines are drawn with 1 MeV spacing, and thin lines correspond to divisions of 0.2 MeV. The energies in MeV relative to the absolute minimum, marked with a black circle, are shown near thick contour lines. 

It is seen from the presented two-dimensional plots that increase of the axial deformation $\beta$ leads to formation of a pre-fission prolate shape.
In the case of $^{189}$Pt a $\gamma$-soft nature of this nucleus belonging to a transitional region between prolate and oblate deformation has been pointed at in \cite{wei-2009}. This can be compared with the behaviour in the presented energy surface in the form of a shallow valley at $\beta \approx 1.5$.

Equilibrium values of the parameters $\beta$ and $\gamma$, calculated using the presented model are shown in table \ref{table1}. In the table, the nuclei with $N$ or $Z$ within $\pm 1$ from the magic number, for which the (b) set of the parameter values was used, are highlighted. In addition, a comparison with the data from the database of quadrupole deformation values \cite{cdfe} of the Center for Photonuclear Experiments Data (CDFE) and theoretically calculated deformation parameters from \cite{moeller-1992,moeller-1995,*moeller-2016,amedee,hilaire-2007,goriely-2013,bruslib} is also shown. The ``FRD(L)M'' column lists the results of the closely related FRDM and FRDLM macro-microscopic models: for axially-asymmetric nuclei the data in this column corresponds to the calculation \cite{moeller-2012} based on the FRDLM model, and for axially-symmetric isotopes (i.e., where $\gamma=0$ or $\gamma=\pi/3$) the revised parameter values from the FRDM(2012) model calculation \cite{moeller-2016} is shown.

\begingroup
\begin{table*}[htb]
\small
\caption{Ground state deformation parameters of $^{50}$Cr---$^{241}$Am. Results of the present work are compared with: (a) "FRD(L)M": the macro-microscopic models FDRM \cite{moeller-2016} and FRDLM \cite{moeller-2012}; (b) BRUSLIB database based on the Hartree-Fock-Bogoliubov method with the extended Skyrme force (HFB 1-29 version) \cite{goriely-2013,bruslib}; (c) AMEDEE database \cite{amedee} based on the Hartree-Fock-Bogoliubov calculations using the D1S Gogny force; (d) deformation parameters based on the experimental data from the CDFE database \cite{cdfe} (if not indicated explicitly, the sign of $\beta_2$ is unknown).}
\label{table1}
\begin{tabular}{r||rrrr|rrr|rr|rr|r}
Isotope & \multicolumn{4}{|c|}{This work} & \multicolumn{3}{|c|}{FRD(L)M \cite{moeller-2016,moeller-2012}} & \multicolumn{2}{|c|}{BRUSLIB \cite{goriely-2013}} & \multicolumn{2}{|c|}{AMEDEE \cite{amedee,delaroche-2010}} & CDFE \cite{cdfe} \\
\rule[-0.5em]{0pt}{1.2em} & $\beta$ & $\gamma$ & $\beta_2$ & $\beta_4$ & $\beta_2$ & $\gamma$ & $\beta_4$ & $\beta_2$ & $\beta_4$ & $\beta_2$ & $\gamma$ & $\beta_2$ \\
\hline
\rule[0pt]{0pt}{1.2em}
 $^{50}$Cr         & 0.28 &  0 & $ 0.282$ & 0.091 & $ 0.194$ & 0  & $ 0.038$ & $0.24 $ & $0.08 $ & $0.240$ & 0.  & $+0.307 \pm 0.075$\\
\magic{$^{52}$Cr}  & 0.15 &  0 & $ 0.155$ & 0.030 & $ 0.000$ & 0  & $ 0.000$ & $0.19 $ & $0.06 $ & $0.000$ & 0.  & $+0.067 \pm 0.02 $\\
 $^{55}$Mn         & 0.23 &  0 & $ 0.231$ & 0.065 & $ 0.172$ & 0  & $ 0.037$ & $-0.18$ & $0.05 $ & $0.25 $ &     & $+0.206 \pm 0.017$\\
\magic{$^{64}$Ni}  & 0.08 & 60 & $-0.085$ & 0.005 & $-0.094$ & 60 & $-0.008$ & $-0.19$ & $0.04 $ & $0.090$ & 60. & $-0.257 \pm 0.141$\\
 $^{68}$Zn         & 0.12 &  0 & $ 0.110$ & 0.012 & $-0.136$ & 60 & $-0.027$ & $-0.22$ & $0.04 $ & $0.140$ & 60. & $ 0.064 \pm 0.015$\\
 $^{68}$Ga         & 0.17 & 34 & $ 0.165$ & 0.027 & $-0.207$ & 45 & $-0.006$ & $-0.23$ & $0.04 $ & $-0.20$ &     & $ 0.022 \pm 0.002$\\
 $^{76}$Ge         & 0.18 &  0 & $ 0.179$ & 0.038 & $ 0.161$ & 0  & $ 0.022$ & $0.21 $ & $0.03 $ & $0.147$ & 0.  & $+0.098 \pm 0.036$\\
 $^{75}$Se         & 0.20 & 60 & $-0.195$ & 0.036 & $-0.238$ & 60 & $-0.023$ & $-0.25$ & $0.04 $ & $-0.05$ &     & $ 0.428 \pm 0.099$\\
 $^{75}$Kr         & 0.38 &  0 & $ 0.406$ & 0.168 & $ 0.402$ & 0  & $-0.010$ & $-0.20$ & $0.01 $ & $-0.15$ &     & $ 0.412 \pm 0.065$\\
\magic{$^{87}$Kr}  & 0.07 & 60 & $-0.062$ & 0.004 & $-0.073$ & 60 & $-0.010$ & $-0.13$ & $0.02 $ & $-0.05$ &     & $-0.102 \pm 0.015$\\
 $^{76}$Rb         & 0.38 &  0 & $ 0.390$ & 0.157 & $ 0.403$ & 0  & $-0.024$ & $-0.20$ & $0.01 $ & $0.50 $ &     & $ 0.482 \pm 0.214$\\
 $^{77}$Rb         & 0.37 &  0 & $ 0.371$ & 0.144 & $ 0.403$ & 0  & $-0.022$ & $-0.23$ & $0.03 $ & $0.00 $ &     & $ 0.441 \pm 0.047$\\
 $^{77}$Sr         & 0.40 &  0 & $ 0.424$ & 0.179 & $ 0.403$ & 0  & $-0.022$ & $-0.19$ & $0.00 $ & $0.50 $ &     & $ 0.481 \pm 0.062$\\
 $^{97}$Mo         & 0.22 & 12 & $ 0.213$ & 0.057 & $ 0.172$ & 32 & $ 0.036$ & $-0.17$ & $0.03 $ & $0.05 $ &     & $ 0.074 \pm 0.031$\\
 $^{98}$Mo         & 0.22 &  0 & $ 0.219$ & 0.051 & $ 0.206$ & 28 & $ 0.017$ & $0.19 $ & $0.04 $ & $0.000$ & 0.  & $+0.089 \pm 0.035$\\
 $^{106}$Cd        & 0.17 &  0 & $ 0.168$ & 0.026 & $ 0.151$ & 0  & $-0.015$ & $0.19 $ & $0.02 $ & $0.145$ & 0.  & $+0.079 \pm 0.027$\\
\magic{$^{110}$In} & 0.10 &  6 & $ 0.113$ & 0.006 & $ 0.097$ & 0  & $-0.021$ & $0.18 $ & $0.02 $ & $0.15 $ &     & $+0.095 \pm 0.01 $\\
\magic{$^{121}$Sn} & 0.02 & 60 & $-0.012$ & 0.002 & $-0.094$ & 60 & $-0.008$ & $0.07 $ & $-0.02$ & $0.00 $ &     & $-0.05  \pm 0.013$\\
 $^{147}$Sm        & 0.20 &  0 & $ 0.192$ & 0.051 & $ 0.140$ & 0  & $ 0.043$ & $0.14 $ & $0.03 $ & $0.10 $ &     & $-0.028 \pm 0.005$\\
 $^{148}$Sm        & 0.20 &  0 & $ 0.201$ & 0.054 & $ 0.172$ & 0  & $ 0.060$ & $0.16 $ & $0.03 $ & $0.167$ & 0.  & $+0.175 \pm 0.061$\\
 $^{149}$Sm        & 0.22 &  0 & $ 0.213$ & 0.062 & $ 0.183$ & 0  & $ 0.062$ & $0.20 $ & $0.09 $ & $0.15 $ &     & $+0.007 \pm 0.003$\\
 $^{150}$Sm        & 0.23 &  0 & $ 0.236$ & 0.071 & $ 0.205$ & 0  & $ 0.066$ & $0.22 $ & $0.09 $ & $0.204$ & 0.  & $+0.226 \pm 0.046$\\
 $^{151}$Sm        & 0.25 &  0 & $ 0.248$ & 0.075 & $ 0.227$ & 0  & $ 0.082$ & $0.27 $ & $0.12 $ & $0.25 $ &     & $+0.093 \pm 0.014$\\
 $^{152}$Sm        & 0.28 &  0 & $ 0.291$ & 0.105 & $ 0.237$ & 0  & $ 0.097$ & $0.28 $ & $0.13 $ & $0.273$ & 0.  & $+0.293 \pm 0.018$\\
 $^{153}$Sm        & 0.30 &  0 & $ 0.301$ & 0.112 & $ 0.259$ & 0  & $ 0.102$ & $0.33 $ & $0.17 $ & $0.30 $ &     & $+0.308 \pm 0.047$\\
 $^{154}$Sm        & 0.30 &  0 & $ 0.311$ & 0.121 & $ 0.270$ & 0  & $ 0.105$ & $0.32 $ & $0.15 $ & $0.347$ & 0.  & $+0.319 \pm 0.023$\\
 $^{160}$Ho        & 0.28 &  0 & $ 0.290$ & 0.104 & $ 0.261$ & 0  & $ 0.051$ & $0.31 $ & $0.09 $ & $0.35 $ &     & $+0.305 \pm 0.03 $\\
 $^{171}$Lu        & 0.32 &  0 & $ 0.325$ & 0.101 & $ 0.287$ & 0  & $-0.030$ & $0.31 $ & $0.05 $ & $0.35 $ &     & $ 0.287 \pm 0.018$\\
 $^{181}$Ta        & 0.23 &  0 & $ 0.245$ & 0.043 & $ 0.255$ & 0  & $-0.076$ & $0.28 $ & $-0.04$ & $0.30 $ &     & $ 0.253 \pm 0.015$\\
 $^{189}$Pt        & 0.17 & 60 & $-0.171$ & 0.029 & $ 0.175$ & 17 & $-0.062$ & $0.21 $ & $-0.05$ & $-0.15$ &     & $-0.186 \pm 0.043$\\
 $^{193}$Hg        & 0.13 & 60 & $-0.137$ & 0.017 & $-0.125$ & 60 & $-0.029$ & $-0.16$ & $0.00 $ & $-0.15$ &     & $-0.114 \pm 0.071$\\
 $^{197}$Au        & 0.12 & 60 & $-0.119$ & 0.010 & $-0.125$ & 60 & $-0.017$ & $-0.15$ & $-0.02$ & $-0.60$ &     & $ 0.096 \pm 0.006$\\
\magic{$^{207}$Bi} & 0.02 & 60 & $-0.014$ & 0.000 & $-0.021$ & 60 & $ 0.000$ & $-0.03$ & $0.00 $ & $0.00 $ &     & $-0.042 \pm 0.001$\\
 $^{229}$Th        & 0.32 & 12 & $ 0.331$ & 0.130 & $ 0.184$ & 0  & $ 0.113$ & $0.21 $ & $0.13 $ & $0.25 $ &     & $+0.31  \pm 0.08 $\\
 $^{241}$Pu        & 0.28 &  0 & $ 0.293$ & 0.107 & $ 0.237$ & 0  & $ 0.086$ & $0.29 $ & $0.13 $ & $0.30 $ &     & $ 0.401 \pm 0.152$\\
 $^{241}$Am        & 0.28 &  0 & $ 0.296$ & 0.106 & $ 0.237$ & 0  & $ 0.086$ & $0.29 $ & $0.14 $ & $0.30 $ &     & $+0.207 \pm 0.014$\\
\end{tabular}

$^{b)}$Nuclei, for which the parameter set (b) was used.
\end{table*}
\endgroup

The agreement with the results from the FRD(L)M and CDFE databases is rather adequate, and it is less satisfactory with the HFB results. However, attention is attracted by the observed disagreement of the theoretical and experimental data for the odd isotopes $^{147,149,151}$Sm, when no such disagreement takes place for the neighboring even isotopes, suggesting that a single nucleon orbiting an even-even core can have a substantial effect on the stability of the ground-state deformation. It is important to note that the CDFE database of deformation parameters was constructed in \cite{cdfe} from the compilation of experimental electric static quadrupole moments \cite{stone-2005}. Experimental measurements using, e.g., the Coulomb excitation technique provide the value of the static quadrupole moments in laboratory frame in a model-independent way \cite{de-boer-1968}, but transformation to the intrinsic coordinate frame is needed to obtain the quadrupole deformation parameter from them. Such transformation in relied on the generalized model's assumption that the intrinsic and rotational degrees of freedom can be considered separable, and could introduce significant systematic errors otherwise, if the amplitude of surface oscillations is larger than or comparable with the static deformation of the nucleus or if the deformation is non-axial (such as $^{68}$Ga).

In order to test the parameter choice, we applied the model to description of the ground-state deformation of the isotopic chains of cadmium and strontium. The isotopes $^{95-132}$Cd and $^{74-106}$Sr, having experimental values of mass in the \cite{wang-2017} database, needed for calculation of the pairing gap $\Delta$, were selected for the calculations. The results of the calculation are shown in table \ref{table2}. The obtained deformations of Cd isotopes increase from zero at the beginning of the $50 < N \le 82$ neutron shell to the maximum value $\beta_2 \approx 0.25$ at its middle and again diminish towards the end. The same tendency shows itself in the results of other models, which differ in the predicted maximum magnitude of deformation. The relatively large negative values of the $\beta_2$ parameter, obtained at the center of the shell in the macro-microscopic calculation \cite{moeller-2012,moeller-2016} are, however, at a certain dissonance with the rest of the results, as well as the tendency for zero deformation for even-even isotopes at $A \le 100$ and $A \ge 120$ in \cite{delaroche-2010}. The results of our calculations for the strontium isotopes are also in a generally good agreement with the bulk of data, obtained by the macro-microscopic model, in HFB calculations using the Gogny force (AMEDEE database), and in the experiment. This is in contrast with the HFB 1-29 model (BRUSLIB database), which predicts oblate deformations for majority of isotopes of Sr.

\begin{turnpage}
\begingroup
\squeezetable
\begin{table*}[htbp]
\caption{Deformation parameters of the Cd and Sr isotopes. Same notation as in table \ref{table1}}
\label{table2}
\renewcommand{\arraystretch}{0.7}
\begin{tabular}{r|rrrr|rrr|rr|rr|r||r|rrrr|rrr|rr|rr|r}
Isotope & \multicolumn{4}{|c|}{This work} & \multicolumn{3}{|c|}{FRD(L)M \cite{moeller-2016,moeller-2012}} & \multicolumn{2}{|c|}{BRUSLIB \cite{goriely-2013}} & \multicolumn{2}{|c|}{AMEDEE \cite{amedee,delaroche-2010}} & CDFE \cite{cdfe} & 
Isotope & \multicolumn{4}{|c|}{This work} & \multicolumn{3}{|c|}{FRD(L)M} & \multicolumn{2}{|c|}{BRUSLIB} & \multicolumn{2}{|c|}{AMEDEE} & CDFE \\
\rule[-0.5em]{0pt}{1.2em} & $\beta$ & $\gamma$ & $\beta_2$ & $\beta_4$ & $\beta_2$ & $\gamma$ & $\beta_4$ & $\beta_2$ & $\beta_4$ & $\beta_2$ & $\gamma$ & $\beta_2$ &
\rule[-0.5em]{0pt}{1.2em} & $\beta$ & $\gamma$ & $\beta_2$ & $\beta_4$ & $\beta_2$ & $\gamma$ & $\beta_4$ & $\beta_2$ & $\beta_4$ & $\beta_2$ & $\gamma$ & $\beta_2$ \\
\hline
\rule[0pt]{0pt}{1.2em}
$^{95}$Cd          & 0.08 &  0 & $ 0.084$ & $-0.011$ & $ 0.053$ & & $ 0.001$ & $0.11 $ & $0.00 $ & 0.10  &    & $                  $ & \magic{$^{131}$Cd} & 0.02   & 60   & $-0.013$ & $-0.000$ & $-0.011$ & & $ 0.000$ & $-0.04$ & $0.01 $ & 0.00    &       & $              $   \\
$^{96}$Cd          & 0.07 &  0 & $ 0.062$ & $ 0.004$ & $-0.021$ & & $ 0.000$ & $0.09 $ & $0.01 $ & 0.000 & 0. & $                  $ & $^{132}$Cd         & 0.03   &  0   & $ 0.031$ & $ 0.002$ & $ 0.000$ & & $ 0.000$ & $0.00 $ & $0.00 $ & 0.000   & 0.    & $              $   \\
\cline{14-26}
\magic{$^{97}$Cd}  & 0.03 &  0 & $ 0.042$ & $-0.004$ & $ 0.043$ & & $-0.011$ & $0.08 $ & $0.00 $ & 0.05  &    & $                  $ & $^{74}$Sr          & $0.42$ &  $0$ & $0.442$  & $0.200$  &  $0.401$ & &  $0.001$ & $-0.23$ &  $0.02$ & $0.170$ & $60.$ &                    \\
\magic{$^{98}$Cd}  & 0.00 &  0 & $ 0.035$ & $ 0.015$ & $-0.021$ & & $ 0.000$ & $-0.03$ & $0.01 $ & 0.000 & 0. & $                  $ & $^{75}$Sr          & $0.40$ &  $0$ & $0.408$  & $0.173$  &  $0.402$ & & $-0.012$ & $-0.24$ &  $0.03$ &  $0.50$ &       &                    \\
\magic{$^{99}$Cd}  & 0.02 & 60 & $-0.013$ & $ 0.000$ & $-0.032$ & & $ 0.000$ & $-0.06$ & $0.01 $ & 0.05  &    & $                  $ & $^{76}$Sr          & $0.42$ &  $0$ & $0.443$  & $0.196$  &  $0.402$ & & $-0.010$ & $-0.24$ &  $0.03$ & $0.503$ &  $0.$ &                    \\
$^{100}$Cd         & 0.10 &  0 & $ 0.094$ & $ 0.012$ & $-0.032$ & & $ 0.000$ & $0.07 $ & $0.01 $ & 0.000 & 0. & $                  $ & $^{77}$Sr          & $0.40$ &  $0$ & $0.424$  & $0.179$  &  $0.403$ & & $-0.022$ & $-0.19$ &  $0.00$ &  $0.50$ &       & $+0.481 \pm 0.062$ \\
$^{101}$Cd         & 0.13 &  0 & $ 0.121$ & $ 0.021$ & $ 0.075$ & & $ 0.002$ & $0.09 $ & $0.01 $ & 0.10  &    & $                  $ & $^{78}$Sr          & $0.40$ &  $0$ & $0.423$  & $0.178$  &  $0.403$ & & $-0.020$ & $-0.22$ &  $0.02$ & $0.000$ &  $0.$ &                    \\
$^{102}$Cd         & 0.15 &  0 & $ 0.146$ & $ 0.026$ & $ 0.107$ & & $ 0.016$ & $0.11 $ & $0.03 $ & 0.118 & 0. & $                  $ & $^{79}$Sr          & $0.38$ &  $0$ & $0.396$  & $0.167$  &  $0.403$ & & $-0.020$ & $-0.25$ &  $0.03$ & $-0.05$ &       & $+0.442 \pm 0.058$ \\
$^{103}$Cd         & 0.15 &  0 & $ 0.144$ & $ 0.028$ & $ 0.118$ & & $ 0.018$ & $0.14 $ & $0.04 $ & 0.15  &    & $-0.184  \pm  0.171$ & $^{80}$Sr          & $0.03$ &  $0$ & $0.029$  & $0.001$  &  $0.403$ & & $-0.020$ & $-0.22$ &  $0.01$ & $0.000$ &  $0.$ &                    \\
$^{104}$Cd         & 0.17 &  0 & $ 0.165$ & $ 0.028$ & $ 0.129$ & & $ 0.007$ & $0.16 $ & $0.04 $ & 0.141 & 0. & $ 0.174  \pm  0.024$ & $^{81}$Sr          & $0.05$ &  $0$ & $0.040$  & $0.002$  &  $0.403$ & & $-0.019$ & $-0.20$ &  $0.01$ & $-0.05$ &       &                    \\
$^{105}$Cd         & 0.17 &  0 & $ 0.164$ & $ 0.030$ & $ 0.140$ & & $-0.004$ & $0.16 $ & $0.04 $ & 0.15  &    & $+0.098  \pm  0.014$ & $^{82}$Sr          & $0.07$ &  $0$ & $0.062$  & $0.005$  &  $0.000$ & &  $0.000$ & $-0.16$ &  $0.01$ & $0.000$ &  $0.$ &                    \\
$^{106}$Cd         & 0.17 &  0 & $ 0.168$ & $ 0.026$ & $ 0.151$ & & $-0.015$ & $0.19 $ & $0.02 $ & 0.145 & 0. & $+0.079  \pm  0.027$ & $^{83}$Sr          & $0.10$ &  $0$ & $0.091$  & $0.015$  &  $0.000$ & &  $0.000$ & $-0.15$ &  $0.01$ &  $0.00$ &       & $+0.192 \pm 0.013$ \\
$^{107}$Cd         & 0.20 &  0 & $ 0.192$ & $ 0.032$ & $ 0.141$ & & $-0.029$ & $0.17 $ & $0.00 $ & 0.15  &    & $+0.153  \pm  0.023$ & $^{84}$Sr          & $0.10$ &  $0$ & $0.095$  & $0.009$  &  $0.011$ & &  $0.000$ & $-0.14$ &  $0.01$ & $0.000$ &  $0.$ &                    \\
$^{108}$Cd         & 0.18 &  0 & $ 0.185$ & $ 0.031$ & $ 0.151$ & & $-0.029$ & $0.21 $ & $0.02 $ & 0.146 & 0. & $+0.126  \pm  0.029$ & $^{85}$Sr          & $0.08$ &  $0$ & $0.081$  & $0.011$  & $-0.053$ & & $-0.011$ & $-0.11$ &  $0.00$ &  $0.05$ &       & $+0.060 \pm 0.006$ \\
$^{109}$Cd         & 0.23 &  0 & $ 0.229$ & $ 0.054$ & $ 0.140$ & & $-0.030$ & $0.21 $ & $0.02 $ & 0.15  &    & $+0.153  \pm  0.023$ & $^{86}$Sr          & $0.07$ &  $0$ & $0.061$  & $0.003$  &  $0.000$ & &  $0.000$ & $-0.11$ &  $0.01$ & $0.000$ &  $0.$ &                    \\
$^{110}$Cd         & 0.27 &  0 & $ 0.273$ & $ 0.080$ & $ 0.152$ & & $-0.041$ & $0.22 $ & $0.02 $ & 0.144 & 0. & $+0.108  \pm  0.022$ & \magic{$^{87}$Sr}  & $0.03$ &  $0$ & $0.044$  & $-0.005$ &  $0.043$ & & $-0.011$ &  $0.09$ & $-0.01$ &  $0.05$ &       & $+0.072 \pm 0.008$ \\
$^{111}$Cd         & 0.25 &  0 & $ 0.257$ & $ 0.072$ & $ 0.162$ & & $-0.040$ & $0.19 $ & $0.00 $ & 0.15  &    & $                  $ & \magic{$^{88}$Sr}  & $0.00$ &  $0$ & $0.027$  & $-0.012$ &  $0.000$ & &  $0.000$ &  $0.08$ &  $0.00$ & $0.000$ &  $0.$ &                    \\
$^{112}$Cd         & 0.27 &  0 & $ 0.275$ & $ 0.077$ & $ 0.174$ & & $-0.039$ & $0.22 $ & $0.01 $ & 0.143 & 0. & $+0.106  \pm  0.027$ & \magic{$^{89}$Sr}  & $0.03$ & $60$ & $-0.029$ & $0.000$  & $-0.032$ & &  $0.000$ & $-0.12$ &  $0.02$ & $-0.05$ &       & $-0.086 \pm 0.007$ \\
$^{113}$Cd         & 0.25 &  0 & $ 0.257$ & $ 0.070$ & $ 0.185$ & & $-0.038$ & $0.22 $ & $0.00 $ & 0.15  &    & $                  $ & $^{90}$Sr          & $0.10$ & $60$ & $-0.097$ & $0.009$  &  $0.000$ & &  $0.000$ & $-0.13$ &  $0.02$ & $0.000$ &  $0.$ &                    \\
$^{114}$Cd         & 0.25 &  0 & $ 0.257$ & $ 0.063$ & $ 0.196$ & & $-0.036$ & $0.23 $ & $0.02 $ & 0.146 & 0. & $+0.094  \pm  0.008$ & $^{91}$Sr          & $0.12$ & $36$ & $0.111$  & $0.014$  &  $0.021$ & &  $0.000$ & $-0.15$ &  $0.02$ &  $0.10$ &       & $+0.014 \pm 0.004$ \\
$^{115}$Cd         & 0.23 &  0 & $ 0.233$ & $ 0.054$ & $-0.228$ & & $-0.026$ & $0.21 $ & $0.00 $ & 0.15  &    & $                  $ & $^{92}$Sr          & $0.18$ & $28$ & $0.186$  & $0.040$  & $-0.135$ & & $-0.005$ & $-0.16$ &  $0.02$ & $0.090$ & $60.$ &                    \\
$^{116}$Cd         & 0.23 &  0 & $ 0.238$ & $ 0.052$ & $-0.238$ & & $-0.034$ & $0.22 $ & $0.00 $ & 0.198 & 0. & $+0.122  \pm  0.038$ & $^{93}$Sr          & $0.27$ &  $0$ & $0.265$  & $0.080$  &  $0.239$ & &  $0.010$ & $-0.18$ &  $0.03$ & $-0.10$ &       & $+0.080 \pm 0.007$ \\
$^{117}$Cd         & 0.22 &  0 & $ 0.225$ & $ 0.041$ & $-0.238$ & & $-0.034$ & $0.22 $ & $0.01 $ & 0.15  &    & $                  $ & $^{94}$Sr          & $0.28$ &  $0$ & $0.291$  & $0.087$  &  $0.263$ & & $-0.010$ & $-0.18$ &  $0.03$ & $0.130$ & $58.$ &                    \\
$^{118}$Cd         & 0.20 &  0 & $ 0.202$ & $ 0.035$ & $-0.238$ & & $-0.035$ & $-0.16$ & $0.02 $ & 0.205 & 0. & $                  $ & $^{95}$Sr          & $0.37$ &  $0$ & $0.394$  & $0.161$  &  $0.308$ & &  $0.010$ & $-0.21$ &  $0.04$ & $-0.15$ &       &                    \\
$^{119}$Cd         & 0.17 &  0 & $ 0.172$ & $ 0.023$ & $ 0.174$ & & $-0.038$ & $0.17 $ & $0.03 $ & 0.15  &    & $                  $ & $^{96}$Sr          & $0.37$ &  $0$ & $0.388$  & $0.163$  &  $0.341$ & &  $0.041$ & $-0.19$ &  $0.03$ & $0.140$ & $60.$ &                    \\
$^{120}$Cd         & 0.17 &  0 & $ 0.165$ & $ 0.023$ & $ 0.140$ & & $-0.030$ & $-0.16$ & $0.01 $ & 0.000 & 0. & $                  $ & $^{97}$Sr          & $0.37$ &  $0$ & $0.373$  & $0.152$  &  $0.352$ & &  $0.044$ & $-0.18$ &  $0.02$ &  $0.50$ &       &                    \\
$^{121}$Cd         & 0.15 &  0 & $ 0.146$ & $ 0.023$ & $ 0.140$ & & $-0.030$ & $0.14 $ & $0.02 $ & 0.10  &    & $                  $ & $^{98}$Sr          & $0.38$ &  $0$ & $0.407$  & $0.180$  &  $0.352$ & &  $0.045$ & $-0.21$ &  $0.03$ & $0.452$ &  $0.$ &                    \\
$^{122}$Cd         & 0.13 &  0 & $ 0.129$ & $ 0.015$ & $-0.104$ & & $-0.042$ & $-0.13$ & $0.00 $ & 0.000 & 0. & $                  $ & $^{99}$Sr          & $0.37$ &  $0$ & $0.389$  & $0.168$  &  $0.352$ & &  $0.045$ & $-0.24$ &  $0.05$ &  $0.45$ &       &                    \\
$^{123}$Cd         & 0.12 &  0 & $ 0.110$ & $ 0.016$ & $-0.104$ & & $-0.042$ & $0.12 $ & $0.03 $ & 0.10  &    & $                  $ & $^{100}$Sr         & $0.37$ &  $0$ & $0.391$  & $0.159$  &  $0.353$ & &  $0.033$ & $-0.25$ &  $0.05$ & $0.448$ &  $0.$ &                    \\
$^{124}$Cd         & 0.10 &  0 & $ 0.095$ & $ 0.008$ & $ 0.000$ & & $ 0.000$ & $-0.10$ & $0.00 $ & 0.000 & 0. & $                  $ & $^{101}$Sr         & $0.37$ &  $0$ & $0.384$  & $0.156$  &  $0.365$ & &  $0.024$ & $-0.25$ &  $0.04$ &  $0.45$ &       &                    \\
$^{125}$Cd         & 0.08 &  0 & $ 0.077$ & $ 0.010$ & $ 0.000$ & & $ 0.000$ & $0.11 $ & $-0.01$ & 0.05  &    & $                  $ & $^{102}$Sr         & $0.37$ &  $0$ & $0.392$  & $0.151$  &  $0.353$ & &  $0.021$ & $-0.26$ &  $0.05$ & $0.445$ &  $0.$ &                    \\
$^{126}$Cd         & 0.08 &  0 & $ 0.077$ & $ 0.004$ & $-0.021$ & & $ 0.012$ & $0.06 $ & $0.01 $ & 0.000 & 0. & $                  $ & $^{103}$Sr         & $0.37$ &  $0$ & $0.390$  & $0.152$  &  $0.355$ & & $-0.005$ &  $0.41$ &  $0.12$ &  $0.45$ &       &                    \\
$^{127}$Cd         & 0.07 &  0 & $ 0.062$ & $-0.007$ & $-0.011$ & & $ 0.000$ & $0.08 $ & $-0.01$ & 0.05  &    & $                  $ & $^{104}$Sr         & $0.35$ &  $0$ & $0.372$  & $0.130$  &  $0.355$ & & $-0.019$ &  $0.41$ &  $0.11$ & $0.438$ &  $0.$ &                    \\
$^{128}$Cd         & 0.03 &  0 & $ 0.028$ & $ 0.001$ & $ 0.000$ & & $ 0.000$ & $0.00 $ & $0.00 $ & 0.000 & 0. & $                  $ & $^{105}$Sr         & $0.33$ &  $0$ & $0.352$  & $0.119$  &  $0.356$ & & $-0.032$ &  $0.41$ &  $0.10$ &  $0.45$ &       &                    \\
\magic{$^{129}$Cd} & 0.02 &  0 & $ 0.025$ & $-0.005$ & $ 0.021$ & & $-0.012$ & $0.05 $ & $-0.01$ & 0.05  &    & $                  $ & $^{106}$Sr         & $0.30$ &  $0$ & $0.314$  & $0.094$  &  $0.345$ & & $-0.049$ &  $0.39$ &  $0.08$ & $0.230$ & $60.$ &                    \\
\magic{$^{130}$Cd} & 0.00 &  0 & $ 0.013$ & $ 0.016$ & $ 0.000$ & & $ 0.000$ & $0.00 $ & $0.00 $ & 0.000 & 0. & $                  $ & & & & & & & & & & & & & \\
\end{tabular}
\end{table*}
\endgroup
\end{turnpage}

\section*{Conclusions}

Deformed one-paticle potential plays an important role in the macro-microscopic approach, since the total energy of occupied single-particle levels contain the fluctuating part of the internal energy of nucleus, which has to be taken into account along with the smooth component, described by the liquid-drop model, for estimation of the equilibrium ground-state deformation. In a sense, the macro-microscopic approach, by separation of the internal energy into the smooth macroscopic and oscillating microscopic components, phenomenologically treats the problem of incorporation of the effects of residual forces for nucleons in a deformed average field.

The model, formulated in this work, presents a different method of correction of the total single-particle energy $E$, allowing to take into account some important aspects of the effect of the residual interaction without leaving the framework of single-particle shell model. This is achieved by consideration of the behaviour of the diffuseness of the nuclear surface (and, correspondingly, behaviour of the realistic one-particle potential), which, as it has been shown, undergoes changes at different value of the deformation, reaching the minimum value near the point of the equilibrium deformation. Therefore, due to decrease of the total energy of nucleons at the surface, the absolute minimum of the energy $E$ is the observed at that point.

As shown by the calculations, only a slight variation ($\le 1\%$) of the diffuseness parameter is necessary for the desired effect. An obvious advantage of the proposed approach is the small number of additional model parameters describing the depedendece of diffuseness on deformation parameters $(\beta,\gamma)$---only three---that were needed for satisfactory description of considered data. All other parameters of the ellipsoidally deformed potential are taken without alteration from the global optical model potential \cite{koning-2003}.

Comparison of the results of calculations performed using the described approach with experimental data and results of macro-microscopic and HFB calculations shows a generally good agreement, suggesting that the described model is not without promise. However, for definitive conclusions larger-scale calculations over a broad range of isotopes are clearly needed. Such calculations are projected to be performed in the future.

The source code of the programs that were used for calculations is made available on the website\footnote{\url{http://bitbucket.org/kstopani2021/deform}}.

\bigskip

Authors would like to dedicate this work to the memory of a distinguished nuclear physicist Prof.~B.~S.~Ishkhanov (1938--2020), a leader who had made a lot for successful work of our group. Without his help and encouragement this, and a great many of other works, would have never seen light. 

\appendix
\section{Estimation of the multipole deformation parameters $\beta_\lambda$}

A common characterization of the nuclear deformation is through the parameters
\begin{equation}
\bt_\lm=\Biggl(\sum_{\mu=-\lm}^\lm |\al_{\lm\mu}|^2\Biggr)^{1/2},
\label{A68}
\end{equation}
where $\al_{\lm\mu}$ are the coefficients of the spherical harmonics expansion of the function $R(\ta\ffi)$ describing the surface of nucleus:
\begin{equation}
R(\ta,\ffi)=R_0\left(
1+\sum_{\lm\ge2,\mu}\al_{\lm\mu}Y^*_{\lm\mu}(\ta\ffi)
\right),
\label{A69}
\end{equation}
where $R_0$ is the radius of the nucleus without deformation and $\lambda$ takes only even integer values $2,4,\dots$ since the considered potential is invariant under spatial inversion.

For prolate spheroidal nuclei $\al_{\lm\mu}=\al_{\lm 0}\,\dl_{\mu 0}$ (where $\dl_{ij}$ is the Kronecker delta). In this case one may rename $\bt_\lm=\al_{\lm 0}$ and refer to it as a signed value. For oblate spheroidal nuclei, to obtain signed $\bt_\lm$ the components of the spherical tensor $\al_{\lm\mu}$ have to be transformed from the coordinate frame $K$, corresponding to the choice of the $(\beta,\gamma)$ plane segment where the symmetry axis of oblate deformation is directed along the $y$ axis, to the coordinate frame $K'$ where the symmetry axis is directed along $z$ using the corresponding rotation matrix \cite{bohr-mottelson-1,edmonds-1957}:
\begin{equation}
\bt_\lm = \al'_{\lm 0} =  \sum_\mu D^{\lm*}_{\mu 0}(\frac{\pi}{2},\frac{\pi}{2},0)\al_{\lm\mu}
\end{equation}

As shown in \cite{bohr-mottelson-2}, the coefficients $\al_{\lm\mu}$ are related to moments
\begin{equation}
M(\lm\mu)=\int\rho({\bf r})r^\lm Y_{\lm\mu}(\ta\ffi)d{\bf r},
\label{A70}
\end{equation}
where the function $\rho(\mathbf{r})$ describes the distribution of the matter density in the ground state of nucleus.

We assume that $\rho(\mathbf{r})$ has, similarly to the described single-particle nuclear potential, the form of a Woods--Saxon distribution (with a corresponding substitution of $R_\text{nucl}$, $a_\text{nucl}$ with $R_0$, $a_0$ in the form factor):
\begin{equation}
\rho({\bf r})=\rho_0\{1+\exp[(r-R(\ta,\ffi))/a(\ta,\ffi)]\}^{-1}.
\label{A71}
\end{equation}

Then, by expanding $\rho(\mathbf{r})$ into power series of $\al_{\lm\mu}$ and keeping only the 1st order terms, one obtains \cite{bohr-mottelson-2}:
\begin{equation}
\rho({\bf r})=\rho_0(r)-R_0\frac{\pt\rho(r)}{\pt r}\sum_{\lm\mu}
\al_{\lm\mu}Y^*_{\lm\mu}(\ta\ffi),
\label{A72}
\end{equation}
where
\begin{equation}
\rho_0(r)=\rho_0\{1+\exp[(r-R_0)/a_0]\}^{-1}.
\label{A73}
\end{equation}

Substituting \eqref{A72} and \eqref{A73} into \eqref{A70} we find that
\begin{equation}
M(\lm\mu)=(\lm+2)R_0\al_{\lm\mu}\int\limits_0^\infty\rho(r)r^{\lm+1}dr.
\label{A74}
\end{equation}
The integral in \eqref{A74} is easily computed if the terms of the order of $\exp(-R_0/a_0)$ are neglected (see, e.g., \cite{bohr-mottelson-1}), and the following expression is obtained:
\begin{equation}
\al_{\lm\mu}=\frac{4\pi}{3}M(\lm\mu)\Biggl/(AR_0^\lm),
\label{A75}
\end{equation}
where $R_0\approx 1.2\,A^{1/3}$ fm.

In order to estimate the parameters $\beta_\lambda$ one has to obtain the moments $M(\lambda\mu)$. We assume that its average value is equal to the sum of average values of $\langle M(\lambda\mu)\rangle_n$ and $\langle M(\lambda\mu)\rangle_p$, corresponding to the neutron and proton systems in the considered Woods--Saxon potential:
\begin{equation}
\la M(\lm\mu)\ra = \la M(\lm\mu)\ra_n+\la M(\lm\mu)\ra_p
\label{A76}
\end{equation}

For nucleons of one kind
\begin{multline}
\la M(\lm\mu)\ra = 2\sum_{k=1}^{N_1-1}\la k+|r^\lm Y_{\lm\mu}
(\ta\ffi)|k+\ra\\
+2\sum_{k=N_1}^{N_2}v^2_k\la k+|r^\lm
Y_{\lm\mu} (\ta\ffi)|k+\ra,
\label{A77}
\end{multline}
where for the case of odd number of nucleons $v^2_{N_F} = 1/2$.

Computation of the matrix elements in \eqref{A76} poses no difficulty, since calculation of the eigenstates $|k\sigma\rangle$ of the Hamiltonian \eqref{35} was performed in the basis of the isotropic harmonic oscillator. It should be also noted that, since conjugate states can differ only by a phase factor, it is always possible to choose it so that ${\la
k+|r^\lm Y_{\lm\mu}(\ta\ffi)|k+\ra=\la k-|r^\lm Y_{\lm\mu}
(\ta\ffi)|k-\ra}$.


%

\end{document}